\RequirePackage{fix-cm}
\documentclass[twocolumn,epjc3]{svjour3}  
\smartqed  
\RequirePackage{graphicx}
\RequirePackage{amsmath}
\usepackage{lineno}
\usepackage{wasysym}
\usepackage{gensymb}
\usepackage{amsmath}

\usepackage{caption}
\usepackage{subcaption}

\RequirePackage[colorlinks,citecolor=blue,urlcolor=blue,linkcolor=blue]{hyperref}
\begin{document}


\title{CUPID-0: the first array of enriched scintillating bolometers for 0$\nu\beta\beta$ decay investigations}

\subtitle{}
\author{O.~Azzolini\thanksref{Legnaro}
\and M.T.~Barrera\thanksref{Legnaro}
\and J.~W.~Beeman\thanksref{LBNL}
\and F.~Bellini\thanksref{Roma,INFNRoma}
\and M.~Beretta\thanksref{MIB,INFNMiB}
\and M.~Biassoni\thanksref{INFNMiB}
\and C.~Brofferio\thanksref{MIB,INFNMiB}
\and C.~Bucci\thanksref{LNGS}
\and L.~Canonica\thanksref{LNGS,MIT,e2}
\and S.~Capelli\thanksref{MIB,INFNMiB}
\and L.~Cardani\thanksref{INFNRoma}
\and P.~Carniti\thanksref{MIB,INFNMiB}
\and N.~Casali\thanksref{INFNRoma}
\and L.~Cassina\thanksref{MIB,INFNMiB}
\and M.~Clemenza\thanksref{MIB,INFNMiB}
\and O.~Cremonesi\thanksref{INFNMiB}
\and A.~Cruciani\thanksref{Roma,INFNRoma}
\and A.~D'Addabbo\thanksref{LNGS}
\and I.~Dafinei\thanksref{INFNRoma}
\and S.~Di~Domizio\thanksref{Genova,INFNGenova}
\and F.~Ferroni\thanksref{Roma,INFNRoma}
\and L.~Gironi\thanksref{MIB,INFNMiB}
\and A.~Giuliani\thanksref{CNRS,DISAT}
\and P.~Gorla\thanksref{LNGS}
\and C.~Gotti\thanksref{MIB,INFNMiB}
\and G.~Keppel\thanksref{Legnaro}
\and M.~Martinez\thanksref{Roma,INFNRoma}
\and S.~Morganti\thanksref{INFNRoma}
\and S.~Nagorny\thanksref{LNGS,GSSI}
\and M.~Nastasi\thanksref{MIB,INFNMiB}
\and S.~Nisi\thanksref{LNGS}
\and C.~Nones\thanksref{CEA}
\and D.~Orlandi\thanksref{LNGS}
\and L.~Pagnanini\thanksref{LNGS,GSSI}
\and M.~Pallavicini\thanksref{Genova,INFNGenova}
\and V.~Palmieri\thanksref{Legnaro}
\and L.~Pattavina\thanksref{LNGS,GSSI}
\and M.~Pavan\thanksref{MIB,INFNMiB}
\and G.~Pessina\thanksref{INFNMiB}
\and V.~Pettinacci\thanksref{Roma,INFNRoma}
\and S.~Pirro\thanksref{LNGS}
\and S.~Pozzi\thanksref{MIB,INFNMiB}
\and E.~Previtali\thanksref{INFNMiB}
\and A.~Puiu\thanksref{MIB,INFNMiB}
\and F.~Reindl\thanksref{INFNRoma,e3}
\and C.~Rusconi\thanksref{LNGS,USC}
\and K.~Sch\"affner\thanksref{LNGS,GSSI}
\and C.~Tomei\thanksref{INFNRoma}
\and M.~Vignati\thanksref{INFNRoma}
\and A.~Zolotarova\thanksref{CEA}
}

\institute{INFN - Laboratori Nazionali di Legnaro, Legnaro (Padova) I-35020 - Italy \label{Legnaro}
\and
Materials Science Division, Lawrence Berkeley National Laboratory, Berkeley, CA 94720 - USA\label{LBNL}
\and
Dipartimento di Fisica, Sapienza Universit\`{a} di Roma, Roma I-00185 - Italy \label{Roma}
\and
INFN - Sezione di Roma, Roma I-00185 - Italy\label{INFNRoma}
\and
Dipartimento di Fisica, Universit\`{a} di Milano-Bicocca, Milano I-20126 - Italy\label{MIB}
\and
INFN - Sezione di Milano Bicocca, Milano I-20126 - Italy\label{INFNMiB}
\and
INFN - Laboratori Nazionali del Gran Sasso, Assergi (L'Aquila) I-67010 - Italy\label{LNGS}
\and
Massachusetts Institute of Technology, Cambridge, MA 02139 - USA\label{MIT}
\and
Dipartimento di Fisica, Universit\`{a} di Genova, Genova I-16146 - Italy\label{Genova}
\and
INFN - Sezione di Genova, Genova I-16146 - Italy\label{INFNGenova}
\and
CSNSM, Univ. Paris-Sud, CNRS/IN2P3, Universit\'e Paris-Saclay, 91405 Orsay - France\label{CNRS}
\and
DiSAT, Universit\`{a} dell'Insubria, Como I-22100 - Italy\label{DISAT}
\and
Gran Sasso Science Institute, 67100, L'Aquila - Italy\label{GSSI}
\and
CEA-Saclay, DSM/IRFU, 91191 Gif-sur-Yvette Cedex -France\label{CEA}
\and
Department of Physics  and Astronomy, University of South Carolina, Columbia, SC 29208 - USA\label{USC}
}

\thankstext{e2}{Present address: Max-Planck-Institut fr Physik, D-80805 Muenchen, Germany. }
\thankstext{e3}{Present address: Institut fur Hochenergiephysik der AW, A-1050 Wien,  Austria. Atominstitut, Technical University Vienna, A-1020 Wien, Austria.}

\date{Received: date / Accepted: date}

\maketitle

\begin{abstract}
The CUPID-0 detector hosted at the Laboratori Nazionali del Gran Sasso, Italy, is the first large array of enriched scintillating cryogenic detectors for the investigation of $^{82}$Se neutrinoless double-beta decay ($0\nu\beta\beta$). CUPID-0 aims at measuring a background index in the region of interest (RoI) for $0\nu\beta\beta$ at the level of 10$^{-3}$~counts/(keV$\cdot$kg$\cdot$y), the lowest value ever measured using cryogenic detectors. This result can be achieved by a state of the art technology for background suppression and thorough protocols and procedures for detector preparation and construction. In this paper, the different phases of the detector design and construction will be presented, from the material selection (for the absorber production) to the new and innovative detector structure. The successful construction of the detector lead to promising detector performance which are here preliminarily discussed. 
\end{abstract}

\section{Introduction}

Scintillating cryogenic detectors are excellent devices for rare events investigations. Their use was first proposed in 1989 for the detection of solar neutrinos~\cite{Gonzalez}, but huge target masses were needed, and the technology was yet not enough mature. Bolometers are nowadays extensively used both for applied physics~\cite{review1} and fundamental physics~\cite{review2}.

One of the main challenge for next generation bolometric experiments is to increase the experimental sensitivity using larger mass detectors with lower background level in the region of interest (RoI). This is the case of CUORE~\cite{CUORE}, the first-ever ton-scale bolometric experiment searching for $0\nu\beta\beta$. CUORE will have a sensitivity of 9$\times$10$^{25}$~y~\cite{Alduino:2017pni} to the observation of $^{130}$Te $0\nu\beta\beta$. Its limitation is given by the expected background index in the RoI which will be 0.01~counts/(keV$\cdot$kg$\cdot$y)~\cite{Alduino:2017qet}, mainly ascribed to $\alpha$-particle interactions on the detector surfaces.

CUPID~\cite{CUPID_1,CUPID_2} (CUORE Upgrade with Particle IDentification) aims at developing the technology of scintillating bolometers for the realization of a next generation $0\nu\beta\beta$ experiment with sensitivity at the level of 10$^{27}$~y, depending on the isotope of interest. This goal establishes some technical challenges, the most relevant one is the operation of a ton of isotope with close-to-zero background level for a ton$\times$year exposure~\cite{IHE}, in the RoI of a few keV around the $\beta\beta$ transition energy.

CUPID-0 (formerly LUCIFER~\cite{LUCIFER}) is the first demonstrator of such technology, operating an array of 26 scintillating bolometers of Zn$^{82}$Se (24 enriched and 2 natural). One of the milestones of CUPID-0 is to demonstrate the feasibility of a close-to-zero background experiment, about 10$^{-3}$~counts/(keV$\cdot$kg$\cdot$y), one order of magnitude better than CUORE. This goal is achieved using scintillating bolometers which allow for particle identification, thus rejecting the $\alpha$-background~\cite{LUCIFER,Beeman:2012ci,Armengaud:2017hit}.\newline

The detector installed in the Hall A of the Gran Sasso Underground Laboratory (LNGS) of INFN, sited in Italy. This unique location ensures an effective shielding against high energy cosmic rays of about 3600~m.w.e..\newline

In this work we describe in details all the procedures for the realization and operation of the CUPID-0 detector, from the production of the fundamental units, the scintillating bolometers, to the processing of the thermal sensors for the signal read-out, but also the surface treatment for the reduction of surface contaminations. A review of the detector operations and performance will also be discussed.\newline

\section{Operation of scintillating bolometers}
A scintillating bolometer is a scintillating crystal absorber which is operated as highly sensitive calorimeter at low temperature. The absorber is kept at cryogenic temperature, few tens of mK, in order to minimize its heat capacity~\footnote{The heat-capacity of a dielectric and diamagnetic crystals scales with the third power of the ratio between its temperature operation and its Debye temperature.}. In these conditions an energy deposit induces a sizeable temperature variation measured by means of a Ge Neutron Transmutation Doped (NTD) thermistor~\cite{NTD}. This induces a perturbation of the crystal lattice which is mediated by phonons, these have energies of the order of few $\mu$eV. Given that the RoI is at few MeV, the statistical fluctuation of the mediators is extremely small, thus allowing for an excellent energy resolution over a wide energy range, at the level of per thousands over few MeV.\newline

When the absorber is also a good scintillator at low temperature, a fraction of the deposited energy in the absorber is converted into a light signal. This can be read out by a suitable light detector (LD) facing the crystal. By means of the read-out of heat and light signal the identification of the type of interacting particle is feasible, thus allowing for the rejection of $\alpha$ particle interactions. Currently, the best choice for light detection in such ultra low background and low temperature environment is the operation of an auxiliary bolometer. CUPID-0 uses Ge absorbers equipped with a thermal sensor similar to the one used in the main absorber.\newline

\begin{figure}[h]
\centering
\includegraphics[width=0.9\columnwidth]{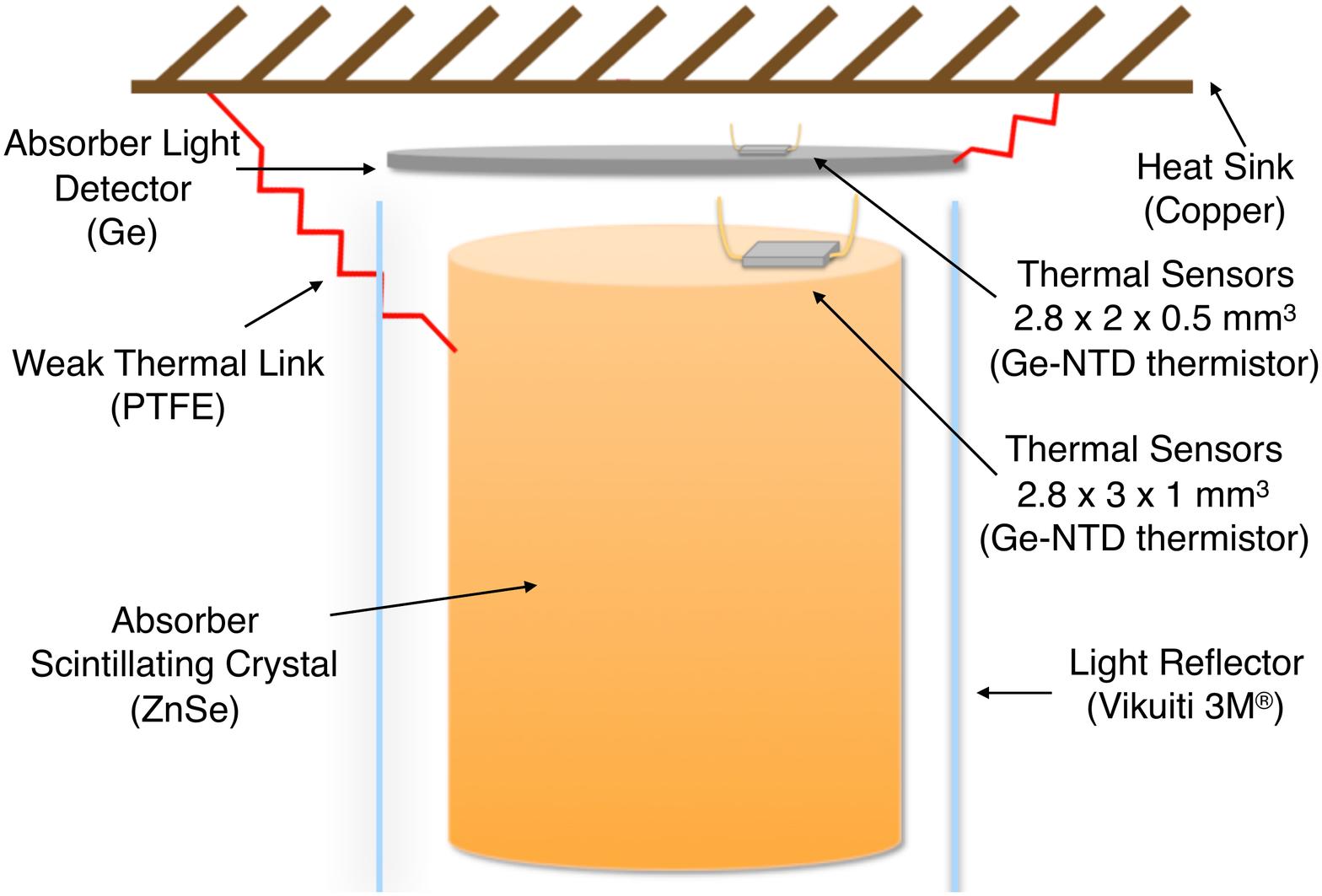}
\caption{\label{fig:bolo}Schematic view of a single module scintillating bolometer.}
\end{figure}

In Fig.~\ref{fig:bolo} a schematic view of a CUPID-0 single module is shown. The ZnSe crystals and the Ge LDs are thermally coupled to the heat bath by means of a copper structure. The two absorbers are held in position by means of PTFE clamps. These act also as weak thermal link to the heat sink and at the same time compensate for the different thermal contractions of the absorbers and of the copper. The difference in mass between the ZnSe, about 450~g, and the LD, about 1~g, requires the design of different thermal sensors which must take into account the different heat capacities. The entire set-up is enclosed in a light reflector which helps in maximizing the light collection efficiency and thus to make the particle identification more efficient.\newline

\section{Design of the CUPID-0 detector array}

The CUPID-0 detector is completely different from any other previously designed bolometric experiment due to its high degree of complexity: large number of channels, extremely compact structure and different absorber dimensions~\footnote{ZnSe crystals and LDs have different dimensions, but also the height of each ZnSe crystal varies.}. This forced the collaboration to the design of a detector structure which had to be reliable, flexible and light at the same time.\newline

CUPID-0 is an array of 26 ZnSe crystals, 24 highly enriched in $^{82}$Se at a level of 95\% and 2 naturals\footnote{The natural isotopic abundance of $^{82}$Se is 8.82\%~\cite{Isotopic_abundance}.}. All the crystals are arranged in a tower like structure and in total there are 5 towers, four containing 5 crystals and one with 6 crystals. The position of the crystals inside the different towers is done in a way such that each tower has about the same weight (height) of about 2~kg (30~cm).

The overall number of $\beta\beta$ nuclei included in the CUPID-0 detector is 3.8$\cdot$10$^{25}$ (natural + enriched crystals).
\newline

\begin{figure}[h]
\centering
\includegraphics[width=1\columnwidth]{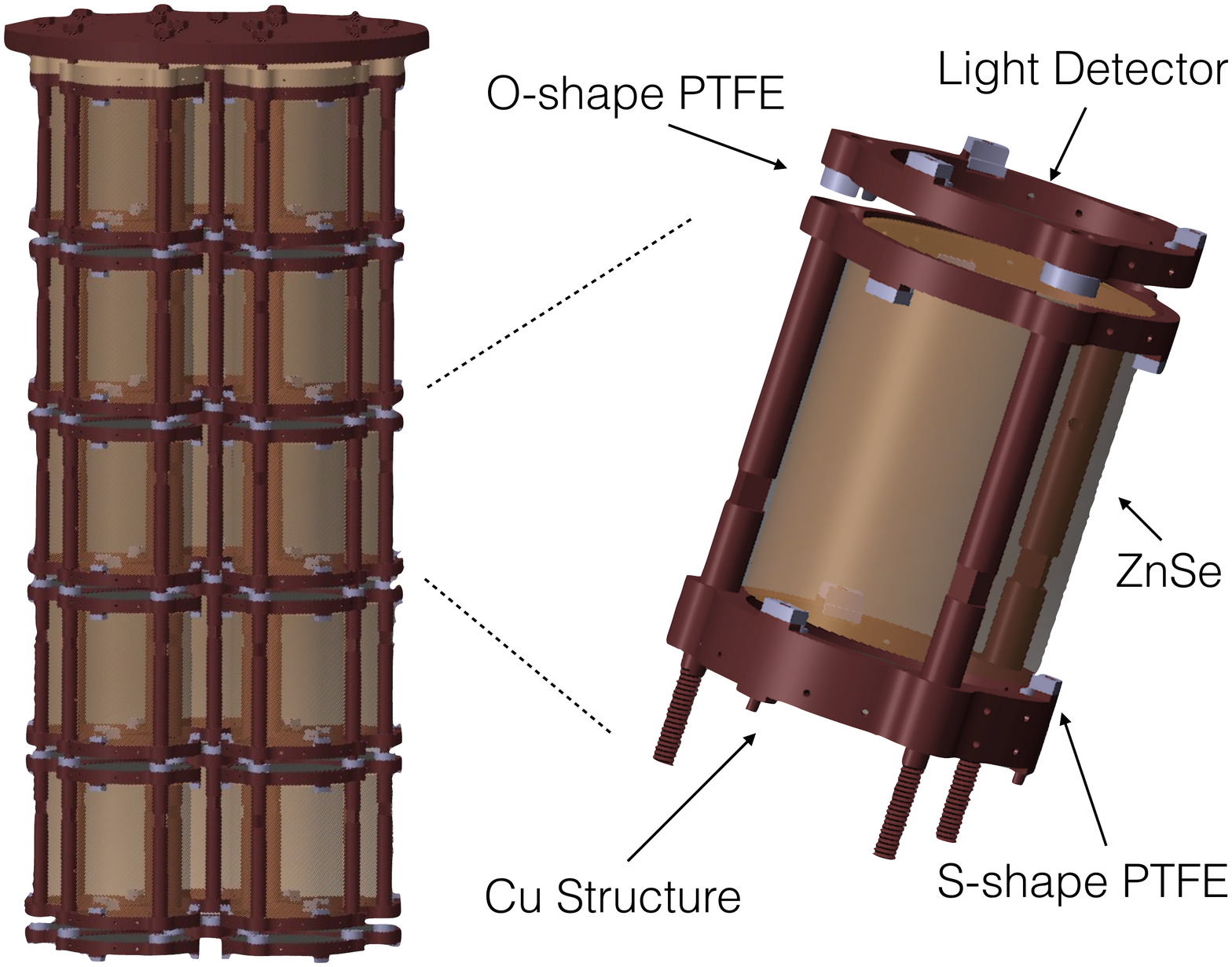}
\caption{\label{fig:rend_tower}Rendering of CUPID-0 detector and its single module.}
\end{figure}

Each tower is about 30~cm tall, the ZnSe crystals are interleaved with Ge-LDs. As it is shown in Fig.~\ref{fig:rend_tower}, each ZnSe is faced to two LDs, one on the top and one on the bottom, so that there is redundancy in the light signal read-out in case of any issue related to their performance.\newline

In order to minimize the amount of passive material next to the detector, which may induce a $\beta/\gamma$ background in the RoI due to its internal radioactive contaminations, the only components selected for the detector construction are: copper, PTFE and light reflectors (VIKUITI-3M). All of the Cu pieces were machined from large Cu chunks from NOSV Cu of Norddeutsche Affinerie AG~\footnote{Now Aurubis AG.}.

The structure is composed by copper frames and columns. The innovative idea for making the simplest possible structure was to use a single copper frame both as ZnSe and LD holder, as it is shown in Fig.~\ref{fig:rend_tower}. In fact the top part of the circular copper component on its upper part holds the crystal, while the bottom part acts on the LD. Progressively on the top of the crystal there is a second frame (like the previous one, but rotate upside-down) where on its top part a new LD is set in place. Using this new type of design for each ZnSe and its adjacent LD only two copper frames are needed. The frames are kept together by means of 3 copper columns, whose lengths is specifically defined by each crystal height. We were able to push down the overall amount of copper in the detector structure and ancillary parts to 22\% of the overall detector mass, an unprecedented value compared to any previous bolometric experiment.

The second material employed for the detector construction is PTFE, which was used to secure in place the two absorbers. This was machined in two sets of pieces: one with a S-shape for the ZnSe and one with a O-shape for the LD. The ZnSe crystals are supported with three S-shaped PTFE on the bottom and three on the top, while the LDs are clamped by three O-shape holders by means of a narrow slit made on the perpendicular axis of the holder. The overall fraction of PTFE in the structure amounts to 0.1\%.

The last component used for the realization of the CUPID-0 detector is the light reflector VIKUITI from 3M. This plastic foil is shaped in cylinders and placed around each crystal, to maximize the light collection efficiency on the two LDs. The reflector completely surrounds the crystals avoiding any line of sight between the absorbers and the closest cryostat radiation shield at 50~mK. This allowed us to prevent the installation of a massive copper shield around the detector at $\sim$10~mK, reducing by a large fraction the copper mass next to the detector.
	
\section{Detector components}

Each component employed for the detector construction was specifically selected for its intrinsic radiopurity. After material screening for each item a dedicated cleaning and purification technique was adopted, the final goal was to mitigate and prevent any recontamination.
This section contains a detailed description of how each detector component was selected and handled before its final installation in CUPID-0.

\subsection{Zn$^{82}$Se crystal absorbers}

For the first time ever large mass Zn$^{82}$Se crystals enriched in $^{82}$Se were grown. The scintillating elements were produced at the Institute for Scintillating Material at Kharkov (Ukraine).
 
The enriched Zn$^{82}$Se crystals were produced starting from highly pure raw materials, namely metal $^{82}$Se and $^{nat}$Zn. The radiopurity of these two metals was investigated at LNGS by means of $\gamma$-spectroscopy on a p-type HP-Ge detector~\cite{Se82}. This detector is characterized by an extremely low intrinsic background that allows the investigation of radioactive contaminations with extremely high sensitivity. In Tab.~\ref{tab:metals} we show the results of the $\gamma$-spectroscopic analysis of the two metals used for the crystal production.

\begin{table}[htp]
\caption{Internal radioactive contamination for 2.5~kg of 96.3\% enriched $^{82}$Se metal beads and for 2.5~kg of  $^{nat}$Zn. Limits are computed at 90\% C.L.. The measurements were carried out on October 2014.} 
\begin{center}
\begin{tabular}{lccc}
\hline\noalign{\smallskip}
Chain & Nuclide  & $^{82}$Se Activity & $^{nat}$Zn Activity\\ 
            & & [$\mu$Bq/kg] & [$\mu$Bq/kg] \\
\noalign{\smallskip}\hline\noalign{\smallskip}
$^{232}$Th & & & \\
 & $^{228}$Ra & $<$~61 & $<$~95 \\
 & $^{228}$Th & $<$~110 & $<$~36\\ 
\noalign{\smallskip}\hline\noalign{\smallskip}
$^{238}$U & & & \\
 & $^{226}$Ra &  $<$~110 & $<$~66\\
 & $^{234}$Th & $<$~6200 & $<$~6200\\
 & $^{234m}$Pa & $<$~3400 & $<$~4700\\
\noalign{\smallskip}\hline\noalign{\smallskip}
$^{235}$U  & $^{235}$U &  $<$~74 & $<$~91\\
\noalign{\smallskip}\hline\noalign{\smallskip}
 & $^{40}$K & $<$~990 & $<$~380\\
\noalign{\smallskip}\hline\noalign{\smallskip}
 & $^{60}$Co & $<$~65 & $<$~36\\
\noalign{\smallskip}\hline\noalign{\smallskip}
 & $^{56}$Co & -- & 80$\pm$20\\
\noalign{\smallskip}\hline\noalign{\smallskip}
 & $^{65}$Zn & -- & 5200$\pm$600\\
\noalign{\smallskip}\hline\noalign{\smallskip}
\end{tabular}
\label{tab:metals} 
\end{center}
\end{table}

As described in~\cite{ZnSe_production}, the synthesis of Zn$^{82}$Se is made in vapour phase by evaporating Zn and enriched $^{82}$Se in Ar atmosphere at 950$^{\degree}$C. The synthesised powder then, after a two stages purification procedure, the first in Ar and the second in H$_2$ atmosphere, is charged inside a high density graphite crucible. All these procedures are performed in Ar flushed disposable glove-boxes, in such a way that the material is never exposed to air so to reduced any possible recontamination.

\begin{figure}[h]
\centering
\includegraphics[width=1\columnwidth]{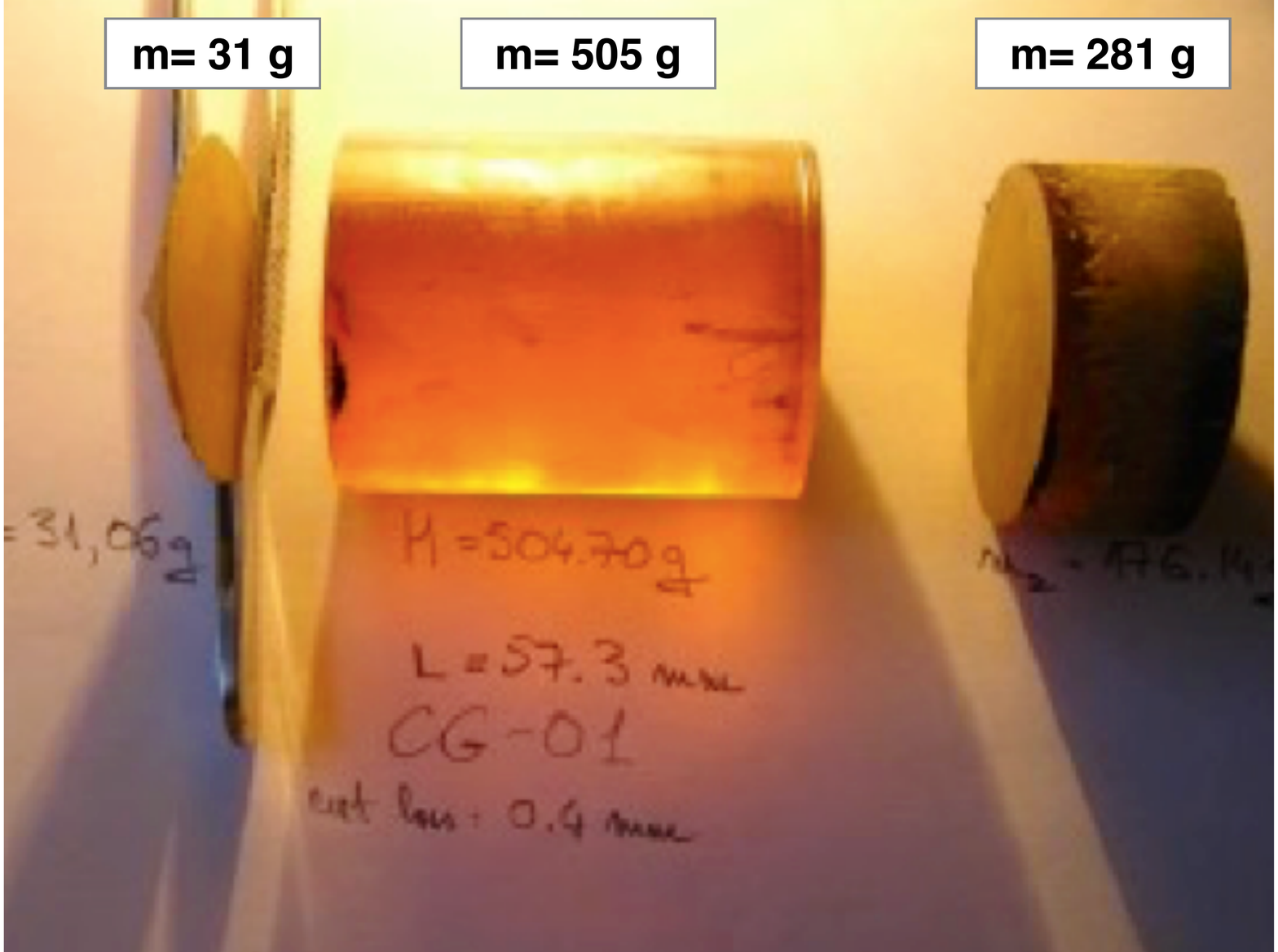}
\caption{\label{fig:znse}Picture of a Zn$^{82}$Se ingot as grown. The two edge of the boule are removed, while the central part is processed for the realization of the final crystal for the CUPID-0 detector.}
\end{figure}

The 1~kg charge is sufficient for the production of a single crystal of 500~g, the rest is all recoverable material which is not included in the final crystal production given its poor crystalline quality, see Fig.~\ref{fig:znse}. The crystal is grown using the Bridgman technique at 1500$^{\degree}$C at about 15~MPa of Ar pressure, with a growing speed of about 1.5~mm/h.

The final crystal is then shaped and optical polished with specific materials and procedures, for further details on the crystal polishing and lapping see~\cite{ZnSe_production}. All these delicate procedures were carried out inside a clean-room where a Radon-abatement system was installed in order to reduce any possible recontamination of the crystals after the polishing.

\begin{figure}[h]
\centering
\includegraphics[width=1\columnwidth]{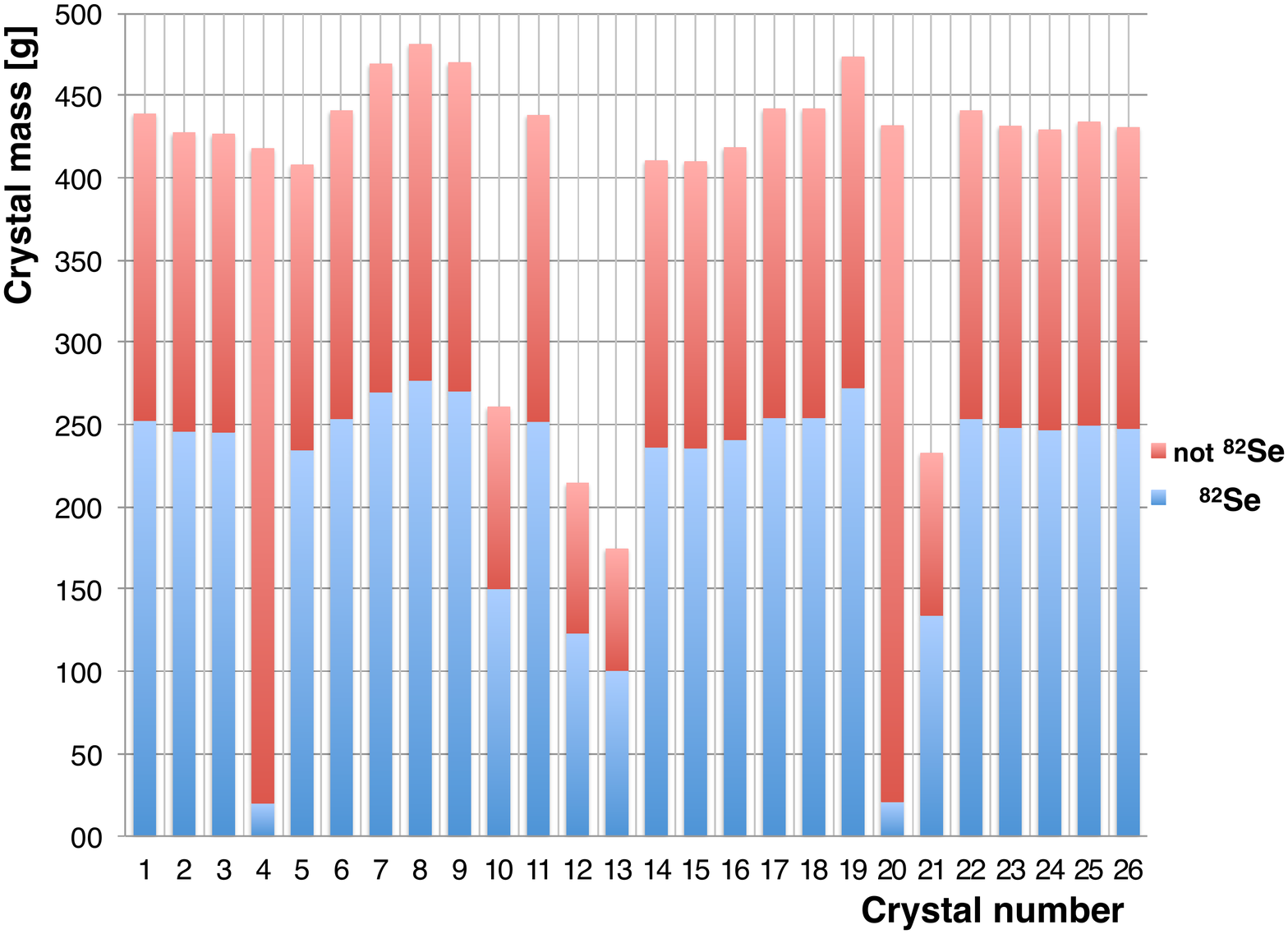}
\caption{Crystal masses. The $^{82}$Se content for each crystal is shown. Crystal number 4 and 20 are crystal with natural Se isotopic abundance.}
\label{fig:fraction}
\end{figure}

In Fig.~\ref{fig:fraction}, the masses of all CUPID-0 crystals and the fraction of the $0\nu\beta\beta$ source are shown. They range from about 150~g to about 500~g, due to the difficult growth conditions. From each ingot, the crystal cuts were performed so that the crystal mass and its quality were maximized. The two crystals with the lowest content of Se are the natural ones.

\subsection{Ge light detectors}
Operating reliable and robust LDs is of paramount importance, since the particle identification relies on the performance of these devices.

The Ge substrate/absorber, purchased at UMICORE Electro-Optic Material (Geel, Belgium), is a double side polished wafer of diameter 44.5~mm and an average thickness of 170~$\mu$m, with an impurity concentration $<2\times$10$^{10}$~atoms/cm$^3$.\newline

All the long-standing experience in the development of cryogenic LDs~\cite{PirroLD} helped in defining the critical issues for an efficient particle identification and rejection, above all: energy resolution and signal amplitude. The former is strongly dependent on the operating conditions of the thermal sensor, as it is thoroughly discussed in~\cite{LD}. The latter mostly depends on the detector design and on the ability to maximize the light detection efficiency. For this reason a dedicated procedure for the enhancement of the light collection efficiency using an anti-reflective coating was developed at CSNSM (Orsay, France). Furthermore, a reflecting foil was employed for focusing on the LD the scintillation light coming from the ZnSe crystal.

\subsubsection{Antireflective coating}
A way to significantly improve the light detector performance is to increase the light collection by minimizing its reflectivity. Such improvement can be achieved by means of a special anti-reflective coating on the side of the detector which is facing the scintillating bolometer.

One of the simplest methods to reduce the reflection is the so called {\it refractive index matching}. In the approximation of normally incident light from a transparent to an absorbing medium, the absorbed fraction can be calculated by the following formula:
\begin{equation}
R=\frac{(n_0-n_1)^2+k^2}{(n_0+n_1)^2+k^2}
\label{reflectance}
\end{equation}
where $n_0$ is the real refraction index of the transparent medium and $n_1(k)$ is the real (imaginary) part of the complex refraction index of the absorbing medium, i.e. germanium. If we consider a simple vacuum-germanium interface, the fraction of absorbed light is only 51\%. On the contrary, if a thin coating layer with refractive index $n_i$, with value between $n_0$ and $n_1$ is placed between the two media, we should evaluate first the $R$ value for the vacuum-coating interface, then for the coating-germanium transition. The optimum value for anti-reflective material to be placed on the Ge detector is a material with $n_i\sim$2.4 which leads to a gain on the light absorption of about 35\% with respect to bare Ge, depending on the wavelength.

The best thickness for the layer can be determined by fulfilling the conditions for an optimal anti-reflective coating in the approximation of single-layer interference. The coating thickness should be $d=\lambda/4$, where $\lambda$ is the wavelength of the incident light. This method works well for monochromatic light sources. In our case, we can take $\lambda$=645~nm, which corresponds to the maximum of the intensity emission for ZnSe scintillation, even though the wavelength distribution is rather broad~\cite{ZnSe_small_light}. The optimal thickness results to be around 65~nm, for $n_i \sim 2.4$.

Several cryogenic tests on anti-reflective coatings were performed at LNGS and at CSNSM during the past years \cite{antireflective}. The best results were achieved with a SiO coating, which has a refractive index of $\sim 2.5$, close to the optimum value previously computed. The increase of the absorbed light fraction was 34\%, in agreement with the expectations. For these reasons a SiO coating was deposited on CUPID-0 LDs.

Forty Ge absorbers were prepared at CSNSM for CUPID-0.
Before the SiO deposition each Ge wafer was previously etched with a mixture of nitric (HNO$_3$ 70\%), acetic (CH$_3$COOH 100\%) and hydrofluoride (HF 40\%) acids in proportion 5:3:3. This mixture is aggressive enough to react with the Ge surface. The thickness of the removed surface layer was  $\approx 10$~$\mu$m with an etching time of about 1~min.
\begin{figure}[h!]
\centering
\includegraphics[width=1\columnwidth]{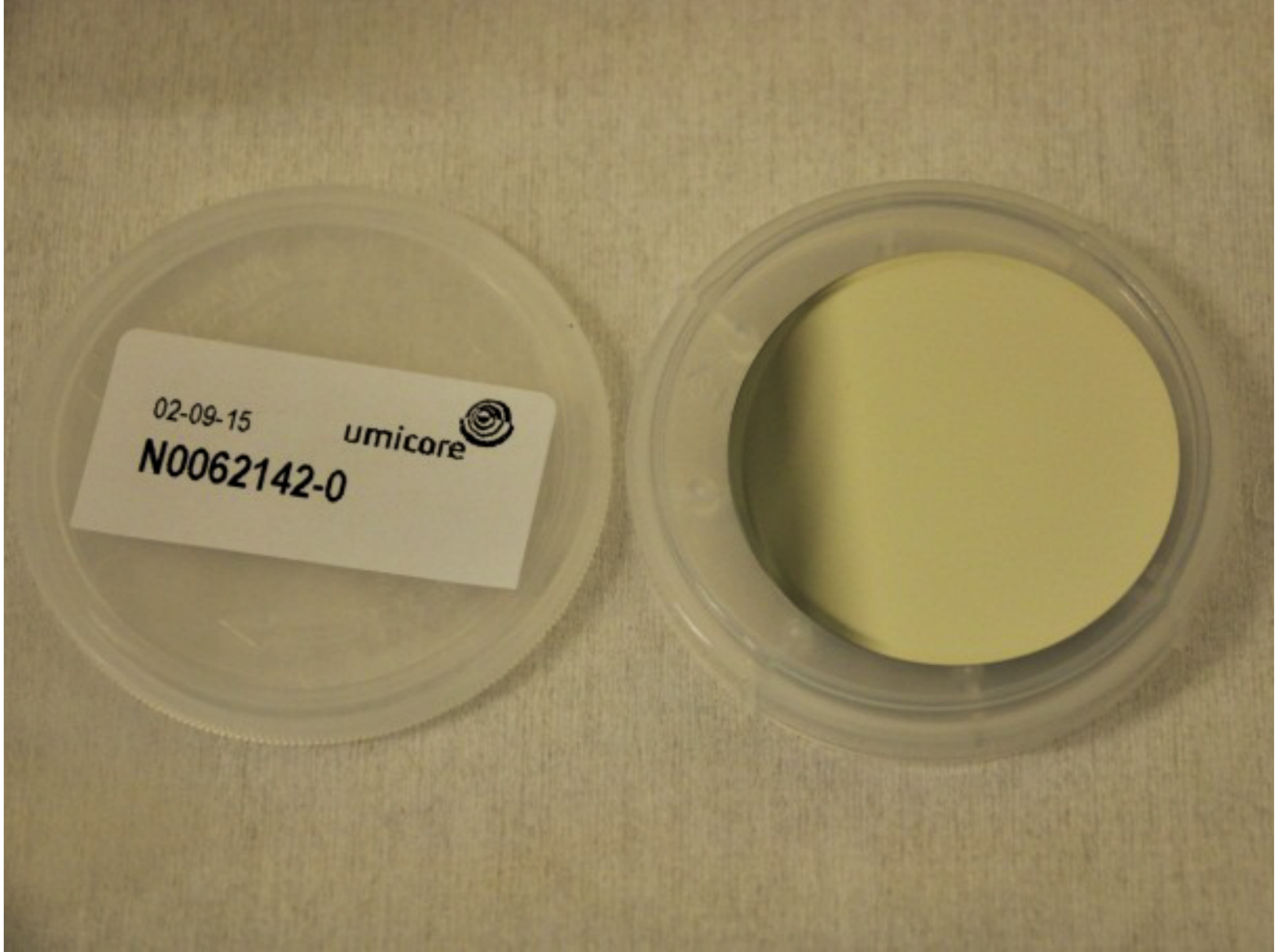}
\caption[BGe]{A high purity Ge slab before applying any procedure.}
\label{fig:BGe}
\end{figure}
After the chemical etching the surface was also treated with an Ar ion bombardment. The gas was ionised with an electron gun, the Ar pressure during the bombardment was 3$\times$10$^{-3}$~mbar.  These procedures remove any possible residual oxides and improve the surface quality for the coating process. The deposition is performed using a tantalum box, where the SiO is heated up to T $\sim 1000$~$\degree C$. The deposition is performed under vacuum: the pressure in the evaporation chamber is P$< 10^{-7}$~mbar.

The evaporation rate was tuned to be in the range of 0.5-1~nm/s. The deposition thickness is controlled with a high precision ($<$~0.1~nm) piezoelectric quartz crystal, its resonance frequency depends on the deposited mass.

\begin{figure}[h]
\centering
\includegraphics[width=1\columnwidth]{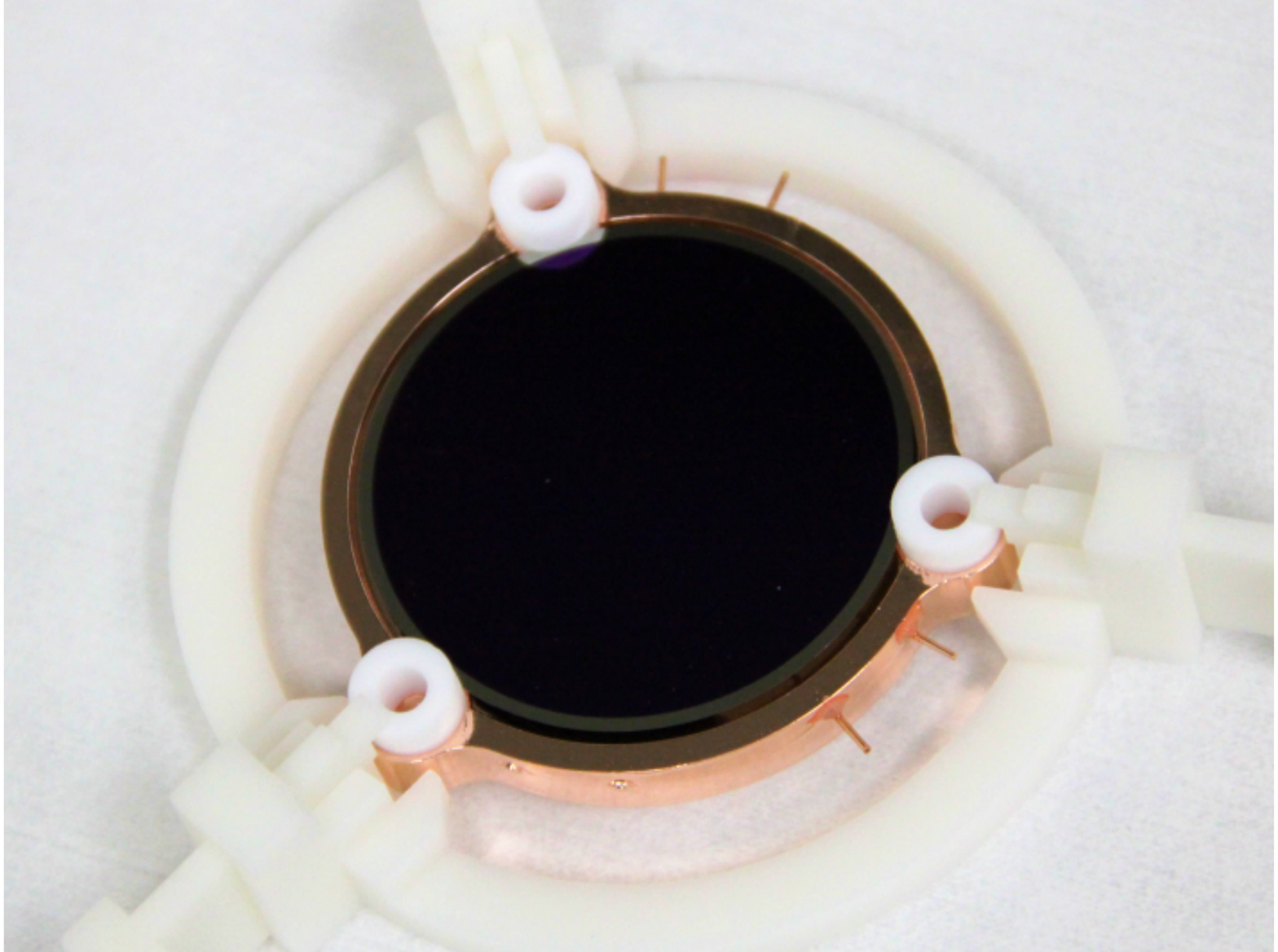}
\caption[BGe]{Light detector before mounting, side with antireflective coating. The coating results in a dark internal circle, while 2~mm on the edge are uncoated.}
\label{fig:SiO}
\end{figure}

The average final thickness of the SiO coatings was between 70-80~nm. This is a good approximation of the value required by the single-layer interference anti-reflective coating, as previously discussed.

\subsection{VIKUITI light reflector}
The light reflector foil installed around the crystal is a VIKUITI multi-layer specular reflector produced by 3M. The foil has been characterized at different temperatures and it ensures a reflectivity greater than 98\% for wavelengths between 400-800~nm~\cite{VIKUITI} over a wide temperature range, from 300~K to 20~K.\newline
The emission spectrum of ZnSe at 10~K has different components, the most intense is at 645~nm~\cite{ZnSe_small_light}, thus ensuring an excellent light collection efficiency.

The reflector radiopurity was investigated with ICP-MS at the LNGS employing an innovative mineralization procedure for the sample preparation~\cite{ICPMS_mineralization}. The measured concentration of the elemental Th and U were 12$\pm$3~ppt and 14$\pm$4~ppt, respectively, see Tab.~\ref{tab:vikuiti}. 

\begin{table}[htp]
\caption{Bulk contaminations of the VIKUITI-3M light reflector. The measurement was carried out at the LNGS by means of a mineralization procedure and an ICPMS analysis. The overall mass of VIKUITI-3M used in CUPID-0 is 17~g. The total activity is 0.8$\pm$0.2~$\mu$Bq and 2.9$\pm$0.7~$\mu$Bq for $^{232}$Th and $^{238}$U, respectively.   } 
\begin{center}
\begin{tabular}{lcc}
\hline\noalign{\smallskip}
Chain & Nuclide  & VIKUITI-3M Activity\\ 
            & & [$\mu$Bq/kg] \\
\noalign{\smallskip}\hline\noalign{\smallskip}
$^{232}$Th & &  \\
 & $^{232}$Th & 49$\pm$12 \\
\noalign{\smallskip}\hline\noalign{\smallskip}
$^{238}$U & &  \\
 & $^{238}$U & 170$\pm$50 \\
 \noalign{\smallskip}\hline\noalign{\smallskip}
\end{tabular}
\label{tab:vikuiti}
\end{center}
\end{table}

\subsection{Ge-NTD thermal sensor}

\begin{figure}[h]
\centering
\includegraphics[width=1\columnwidth]{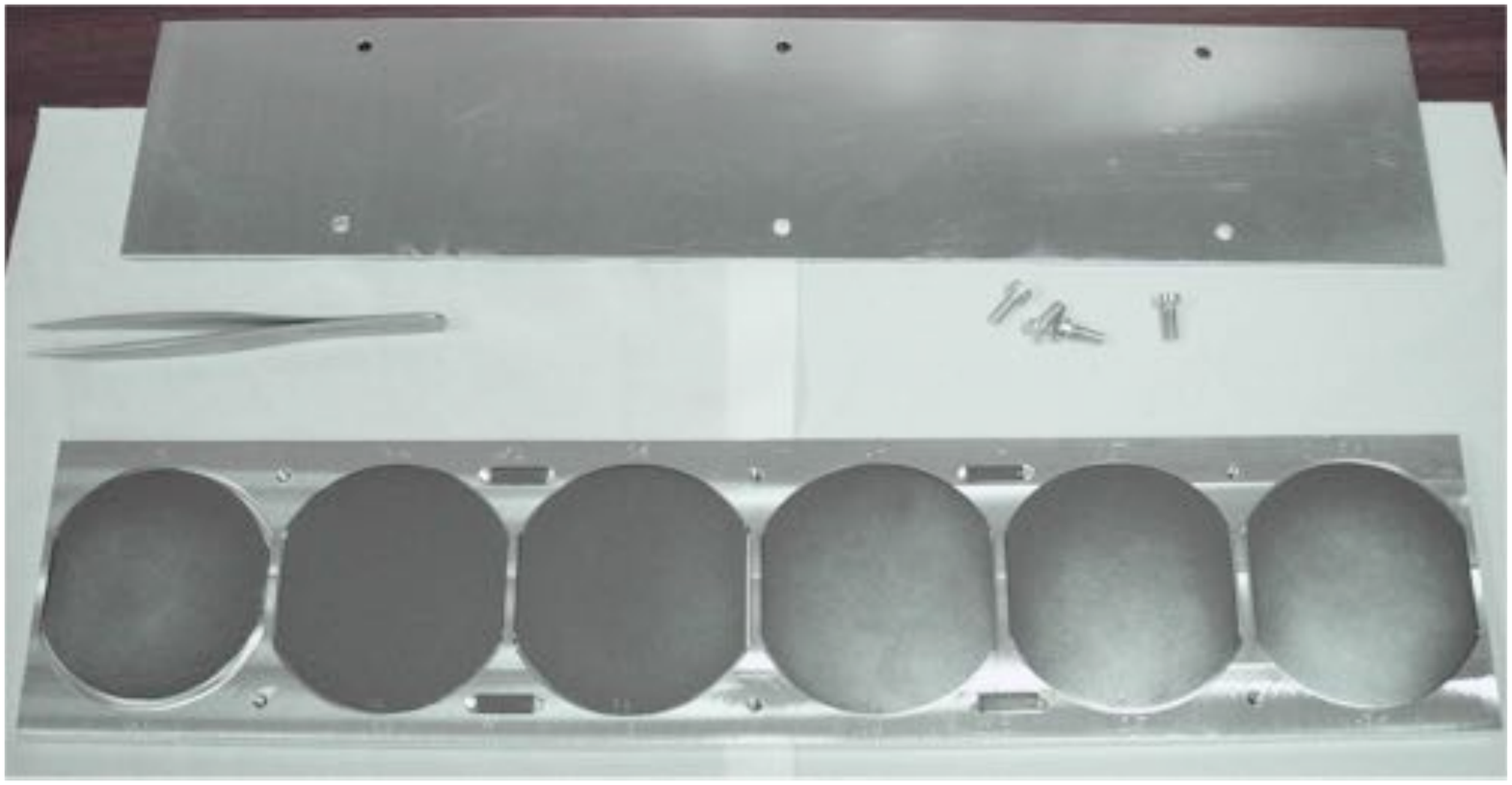}
\caption{Six out of 9 Ge wafers, $\oslash=$65.5~mm and thickness 3.25~mm, installed in an Al holder before neutron irradiation at the MIT Nuclear Reactor Laboratory (MA, USA).}
\label{fig:irr}
\end{figure}

The CUPID-0 thermal sensor production started on March 2012. Nine wafers of high-purity Ge, $\oslash=$~65.5~mm and thickness 3.25~mm, were irradiated at the MIT Nuclear Reactor Laboratory, Boston (MA, USA), see Fig.~\ref{fig:irr}. The exposure to high neutron fluxes is needed in order to uniformly dope the wafers. The required dopants concentration to enable the operation of the Ge as thermal sensor is at level of 10$^{16}$~atom/cm$^3$. Such high and uniform doping level will take the semiconductor close to the metal-to-insulator region. The electrical conductivity of these heavily doped semiconductors has an exponential dependence on the temperature, making these sensors the most suitable technology for our purposes~\cite{NTD}:
\begin{equation}
\rho(T) = \rho_0 \, e^{({T_0/T})^{0.5}},
\end{equation}
where $T_0$ depends on the Ge-NTD doping level and $\rho_0$ on the doping level and on the sensor geometry.
As a consequence, a fluctuation of the doping level will strongly affect the sensitivity of the thermometer, the target value are $T_0$=4.2~K and $\rho_0$=1.5~$\Omega$, having a sensor sensitivity of about 1~M$\Omega$/$\mu$K.

An accurate and precise measurement of the flux and neutron dose to which the Ge wafers are exposed is of paramount importance for the success of the experiment, these values will depend the detector response and resolution. The entire neutron dose should not exceed 2\% of the nominal value, which is estimated to be at the level of 4$\times$10$^{18}$~n/cm$^3$. \newline

To achieve such high accuracy and precision nominal neutron dose, we decided to irradiate: 3 wafers at $\pm$7\% of the target dose, 2 wafers at $\pm$3\%, 1 wafer at +2\% and 3 wafers at the nominal value. This choice was driven by the fact that the Nuclear Reactor facility could ensure dose within 5\% of the target value. At last, a final fine tuning of the neutron dose was carried out at the LENA Nuclear Research Reactor Laboratory (Pavia, Italy), where the most intense neutron flux is about 3 orders of magnitude lower than the MIT one, thus allowing a more accurate neutron irradiation. A detailed mapping of the LENA reactor neutron fluxes was performed~\cite{TRIGA_cold,TRIGA_flux} before irradiating the sensors, this allowed us to select the irradiation channel more suitable for our purposes, thus with a ratio thermal/fast neutron flux of about 20.\newline

\begin{figure}[h]
\centering
\includegraphics[width=1\columnwidth]{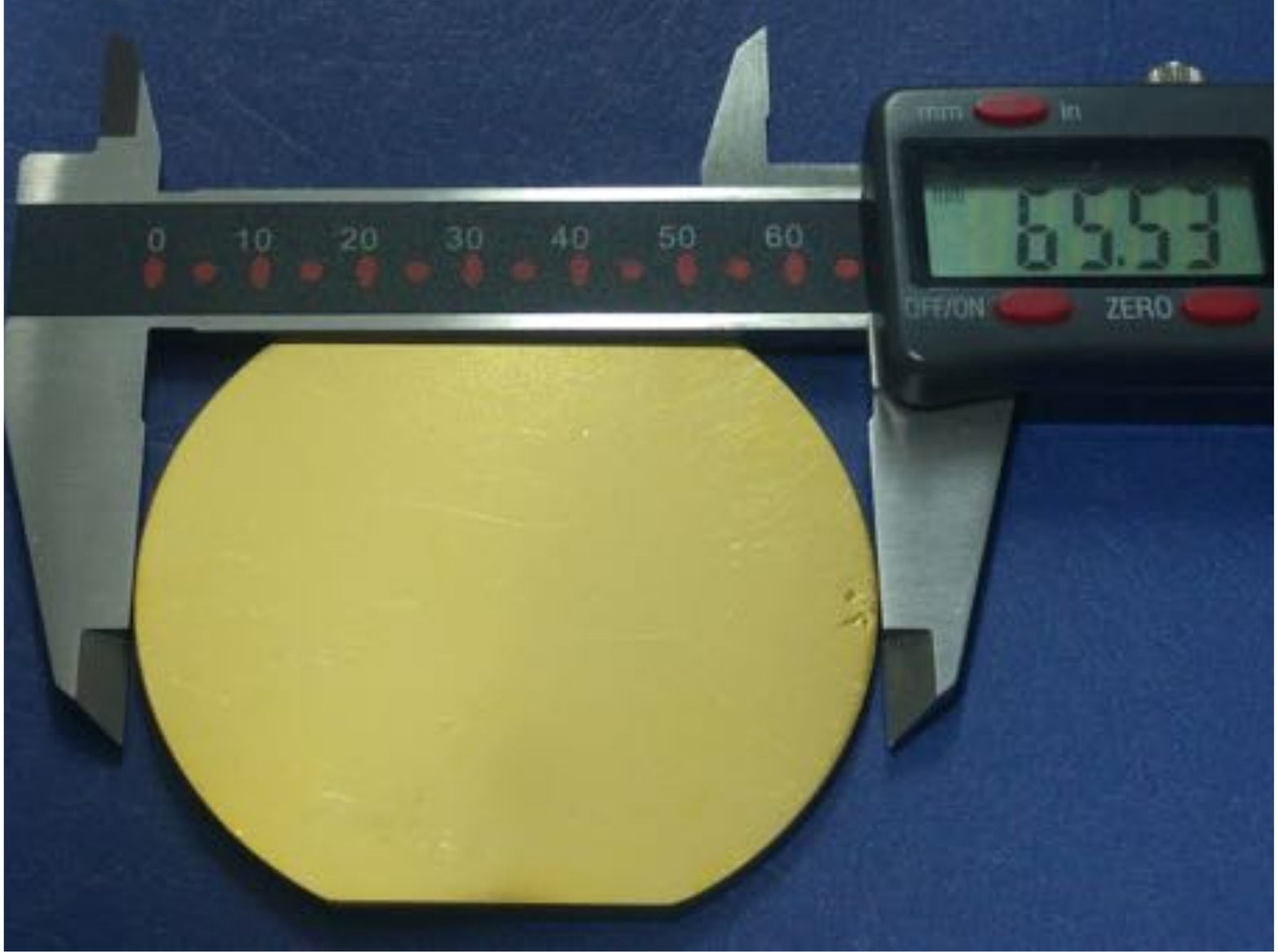}
\caption{Doped Ge wafer with a 4000~$\AA$ gold deposition. The Au on the two wide area of the wafers serve as Ohmic contacts for reading-out the sensor.}
\label{fig:gold}
\end{figure}

\begin{figure}[h]
\centering
\includegraphics[width=1\columnwidth]{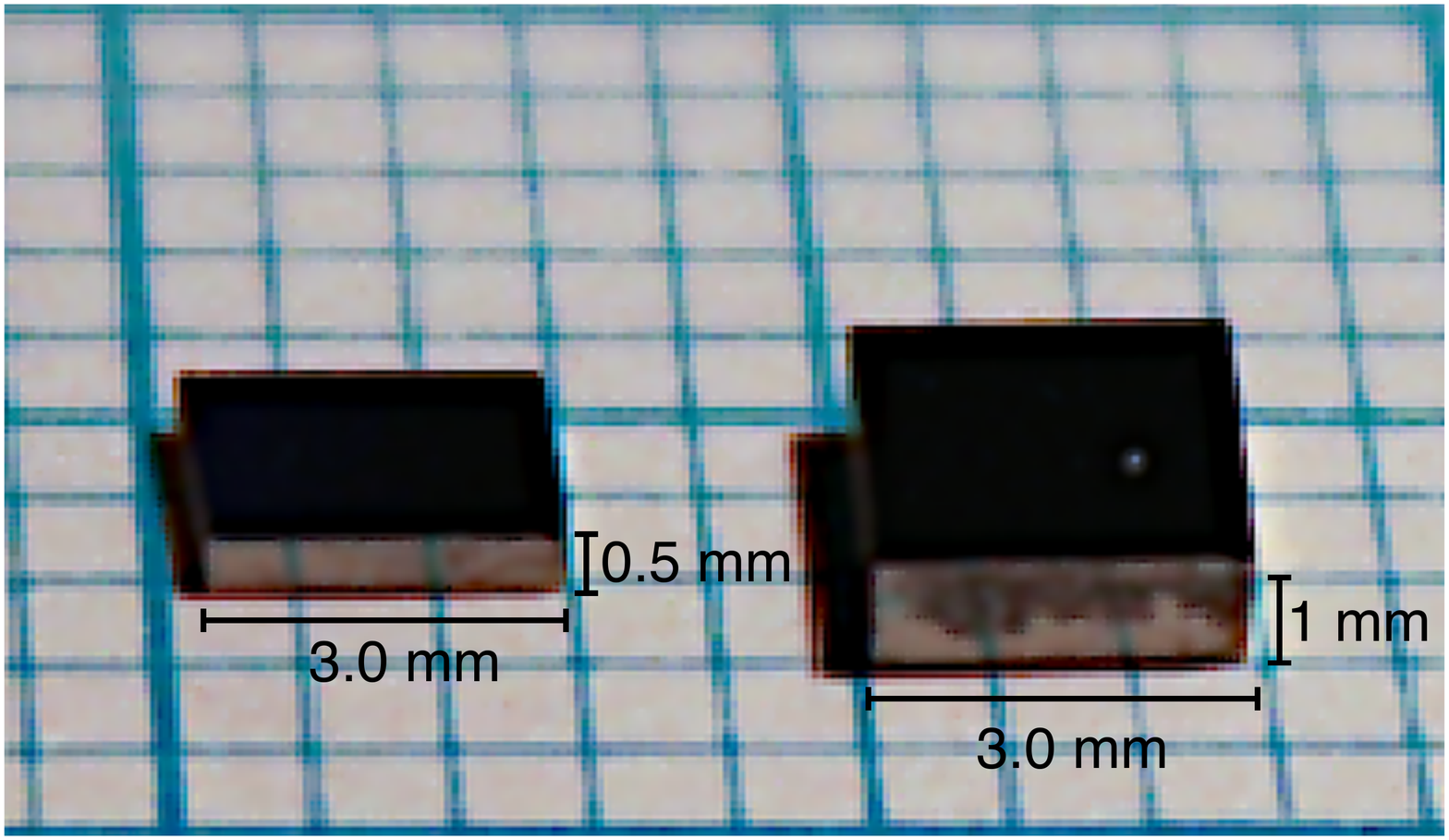}
\caption{Ge Neutron Transmutation Doped sensors for ZnSe (right) and light detector (left). The dimensions for the two sensors are 3$\times$2.8$\times$1~mm$^3$ and  2$\times$2.8$\times$0.5~mm$^3$, respectively }
\label{fig:NTD}
\end{figure}

On the two wide area sides of the Ge wafer a 4000~$\AA$ gold layer is deposited, which serves as sensor Ohmic contacts, see Fig.~\ref{fig:gold}. After the deposition the wafer is diced in sensors of the desired size. The final sensor dimensions for the ZnSe and LDs are 3$\times$2.8$\times$1~mm$^3$ and 2$\times$2.8$\times$0.5~mm$^3$, respectively, see Fig.~\ref{fig:NTD}, where 2.8~mm is the distance between the two golden pads.

The sensors before their coupling to the absorbers are equipped with gold wires. A dedicated ball-bonding machine was used for the bonding of gold wires of 25~$\mu$m of diameter on the golden ohmic contacts.

\subsubsection{Sensors gluing}
The mechanical and thermal coupling of the sensors to the crystal absorbers (referred as gluing) is a well-known concern in the construction of bolometers, because it influences the quality of the detector performance. In particular, the R\&D towards the Cuoricino experiment~\cite{Qino} established stringent requirements to this delicate process, involving the geometry of the glue interface between sensor and absorber, the selection of glue, and the environmental conditions (temperature and humidity) in which the operation has to be performed.

\begin{figure}[h]
\centering
\includegraphics[width=1.1\columnwidth]{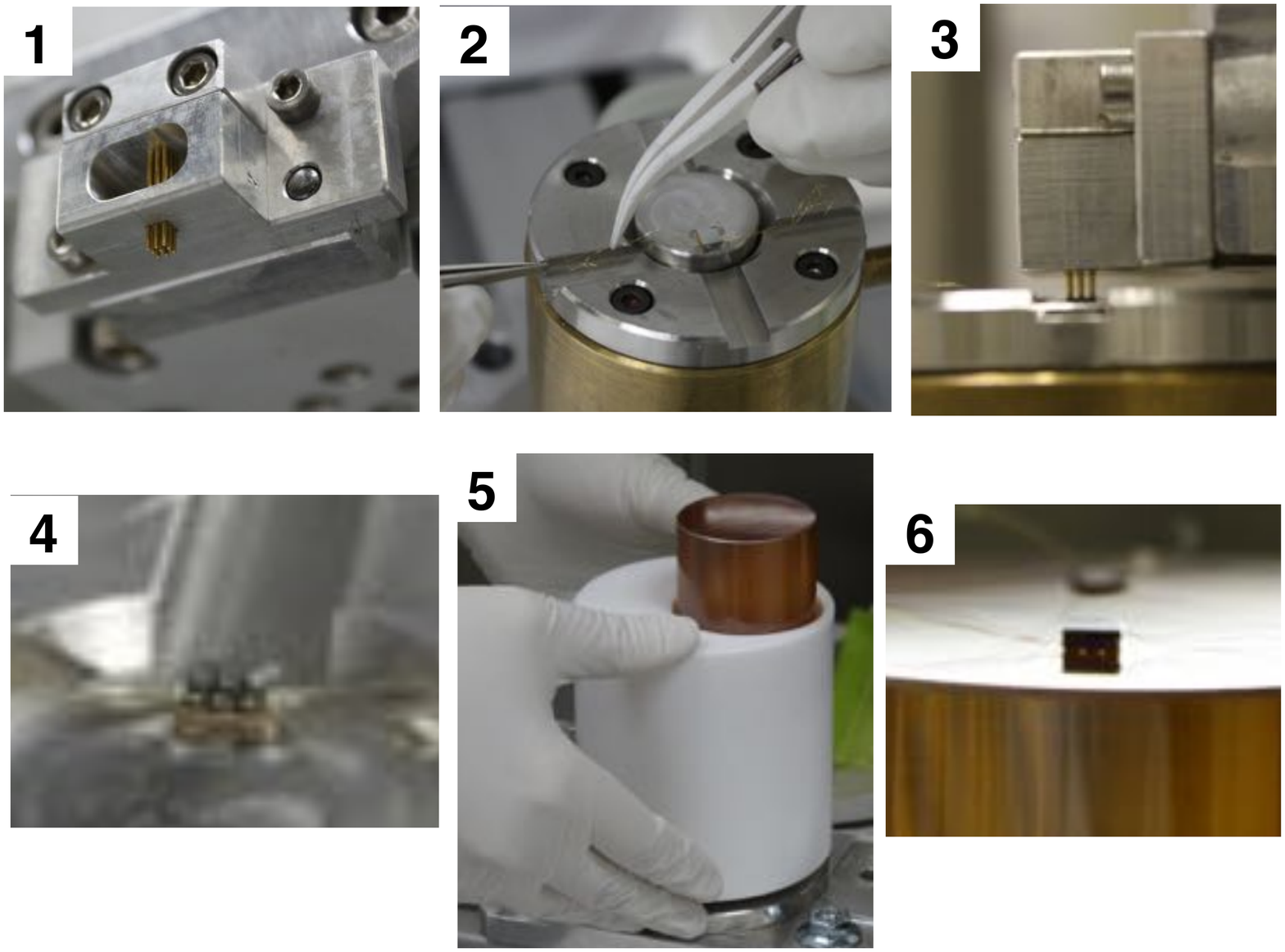}
\caption{Sensor gluing: 1) Matrix of spring-loaded tips used for the 3$\times$2.8$\times$1~mm$^3$ NTDs. 2) Placing NTD on the positioning device and held by vacuum. 3) Spring-loaded tip matrix is lowered on the NTD after dipping in the mixed glue. 4) Example of glue dots on a  2$\times$2.8$\times$0.5~mm$^3$ NTD. 5) A completed gluing, where the array of glue dot is visible between the ZnSe crystal and the NTD.}
\label{fig:dots}
\end{figure}

According to these constraints, it is preferable to deposit the glue in a matrix of dots, in order to compensate for the different thermal contractions at low temperatures. The most suitable number of dots for a 2.8$\times$3~mm$^2$ surface Ge-NTD is 9. These dots must have a diameter of 0.7~mm, while their height is determined by imposing a 0.05~mm gap between the crystal and the absorber, see Fig.~\ref{fig:dots}.

Moreover, the glue is deposited on the sensor surface instead of on the crystal, because the former is easier to reprocess in case of gluing failure, while the cleaning of the crystal surface may require an entire crystal surface polishing treatment.

The glue used for the process must have a density high enough to avoid the merging of the dots after their deposition. Besides that, it has to work at cryogenics temperatures and through several thermal cycles and it must fulfil the radiopurity constraints required by the experiment.
The selected glue is the bi-component epoxy Araldite Rapid by Huntsman Advanced Materials~\cite{colla}. It has a viscosity of 30~Pa$\cdot$s, a very quick pot life of about 3~min, but also a very short curing time of about 1~h and low radioactivity, less than 8.9$\cdot$10$^{-4}$~Bq/kg for $^{232}$Th and 1.0$\cdot$10$^{-2}$~Bq/kg for $^{238}$U~\cite{CUORE-0_detector}.

To improve the reproducibility of the detector performance, a R\&D was carried on to develop a semi-automated system for the sensor-to-absorber coupling of the CUPID-0 bolometers, similar to the one developed for the CUORE experiment~\cite{CUORE_exp}. The CUPID-0 gluing system was developed starting from the previously acquired knowledge~\cite{Crusca}. The result was a semi-automated system in which the dots are performed through a matrix of spring-loaded tips applied to a x-z Cartesian robot, see Fig.~\ref{fig:dots}-1.
The introduction of automated elements reduces the variability induced by manual work and gives the advantage of precise timing, which is useful considering the short time window of epoxy life-time.

The gluing procedure is divided in three steps: the tool preparation, the glue dispensing and the crystal deposition. Among the three, only the second one is automated. Firstly, the Cartesian robot is equipped with the correct spring-loaded tip matrix according to the kind of Ge-NTD to be glued (a nine tip matrix for the 2.8$\times$3$\times$1~mm$^3$ Ge-NTDs for ZnSe crystals and a six tip matrix for the 2.8$\times$2$\times$0.5~mm$^3$ Ge-NTDs for LDs). Then the sensor is placed on a positioning device on the Cartesian robot, where it is held in place by vacuum, see Fig.~\ref{fig:dots}-2. In parallel, the crystal is prepared in a dedicated PTFE holder that will be inserted onto the sensor positioning device after the glue handling phase. This begins with the mixing of the two epoxy components through a dispensing gun provided with a disposable static mixer; the mixed glue is poured and then levelled in a small PTFE container placed on the Cartesian robot, see Fig.~\ref{fig:dots}. The x-z arm dips the tip matrix in the glue container and then presses it onto the sensor surface to deploy the glue dots, see Fig.~\ref{fig:dots}-3. The correct size of the single glue dot is determined by choosing a proper diameter of the tip (0.53~mm) in combination with the depth of the glue container (0.70~mm). The fact that the tips are spring-loaded ensures a uniform collection/deposition of the glue by each tip, see Fig.~\ref{fig:dots}-4.
 
Finally, the crystal is lowered on the Ge-NTD thanks to the PTFE crystal holder in which it was previously hosted, see Fig.~\ref{fig:dots}-5, that ensures a gap of 0.05~mm between the crystal and Ge-NTD surfaces. This, together with the fact that the sensor is held by vacuum for all the glue curing time prevents the dots to merge in a layer, preserving the shape of the dot matrix, see Fig.~\ref{fig:dots}-6.

All the absorbers are also equipped with a Si resistors which are used for injecting fixed amount of energy in the crystals~\cite{Si_heater}. These produces signal similar to the one induced by particle interaction, and they are used for the correction of the detector gain drift caused by the continuous cooling down of the experimental apparatus.

The gluing activity of CUPID-0 was performed inside a Radon-free cleanroom to ensure low radioactive conditions and a very stable environment in terms of temperature and humidity, a mandatory conditions since these parameters influence the glue intrinsic property (especially viscosity).

 \subsection{Ultra-high clean copper}
 
The CUPID-0 detector structure is mainly composed by copper, it makes about 22\% of the overall detector mass. Minimizing the concentration of radioactive impurities, especially from the surface of the detector structure is important for suppressing possible high-energy $\beta/\gamma$ background sources. For this reason a dedicated cleaning procedure was developed for the abatement of surface contamination in copper, similar to the one described in~\cite{TTT}.

The frames that hold the crystals are made of electrolytic tough pitch copper, also known as NOSV copper. Some impurities, including decay products of $^{232}$Th and $^{238}$U decay chain can be accumulate on the material surface as a consequence of an exposure to an uncontrolled atmosphere. Additional impurities are also deposited during several mechanical machining steps needed to produce the copper holders. The concentration of contaminant in the material is usually modelled as a gradient from the first external layer to the inner bulk, caused by diffusion. The radioactive contaminants of $^{232}$Th and $^{238}$U are usually present on copper surfaces to a depth of about 20~$\mu$m~\cite{sticking}.

All the NOSV Cu components were cleaned within a period of 6~months. The cleaning protocol consists of five general macro steps divided in sub-steps for a total of 61 single processes, the time required for cleaning one set was about ten~days. The total number of pieces cleaned was 268: 78 columns (with 26 different lengths), 70 frames, one large copper plate for the tower installation, and several spare parts.

 \subsubsection{Cleaning process protocol to reduce the radioactive contamination levels in Cu components}

The cleaning procedure, developed at the Legnaro National Laboratories (LNL) of INFN consists on a sequence of the successive treatments: Tumbling, Electropolishing, Chemical etching and Magnetron plasma etching (T+E+C+M). The storage of the Cu parts after each cleaning step was performed in a clean-room to avoid possible re-contamination of the surface. 

\begin{itemize} 
\item{\underline{Pre-cleaning process}}: the pre-cleaning is performed for removing any lubricant residues deposited on the copper surfaces and it directly affects the efficiency of the electropolishing process. It is performed wiping the Cu surface using specific wipes and a sequence of three different solvents: tetrachlorethylene to solve organic materials, acetone to degrease and remove tetrachloroethylene, and ethyl alcohol to dissolve the acetone from the Cu surface. Later, the copper pieces are cleaned in ultrasound baths (33~kHz) at 40$^{\degree}$C for 10 minutes with deionized water and NGL 17.40 P.SP powder soap. Right after the bath, the copper pieces are dried with alcohol and nitrogen taking precautions to avoid re-contamination.\newline

\item{\underline{Chemical etching pre-electrochemical process}}: the Cu surfaces must be prepared for the electropolishing process. Tumbling was used for CUPID-0 pieces, except for some delicate parts that cannot undergo the tumbling process, due to the high precision machining and small holes (less than 1~mm). In this case the tumbling step cleaning protocol is substituted by an ammonium persulfate chemical treatment with a concentration of 20~g/L for 2~hours.\newline

\item{\underline{Electrochemical process}}: the electropolishing can remove surface layer up to 50~$\mu$m for the frames and 100~$\mu$m for the other parts. To avoid the removal of Cu in specific area of the detector components such as the threads, PTFE protections were used. The Cu pieces are placed inside an electrochemical solution of 40\% of butanol - 60\% of phosphoric acid, following a designed specific anode Ð cathode configuration for each type of Cu piece. During the process, the Cu surface quality is controlled. After the electrochemical treatment, the residual electro-polishing solution is removed with ultrasonic cleaning.\newline

\item{\underline{Chemical etching process and Passivation}}: the chemical etching is applied to reduce the radioactive contaminants from the areas screened with the PTFE protections. The erosion rate is about 2~$\mu$m/min.
This chemical etching is performed using deionized water heated at~72$\pm$4$^{\degree}$C with a recipe of sulfamic acid, ammonium citrate in powder form adding hydrogen peroxide and butanol in liquid form, named SUBU. The copper pieces are fixed to a sample holder through a copper wire and drawn in the SUBU solution for 5~min, the pieces rotate to enhance the reaction rate. After the SUBU process, the copper pieces are passivated in sulfamic acid solution at a concentration of 20~g/L for 5~minutes and cleaned in ultrasonic bath to finally be dried and packed.\newline

\item{\underline{Plasma cleaning}}: the plasma cleaning constitutes an important step in the cleaning protocol. It is carried out in a vacuum system and it is the last phase before the assembly of the detector. The erosion rate during plasma cleaning is about 1$\mu$m/h. The plasma cleaning is a process based on DC magnetron sputtering technique which consists in the erosion of a target (copper pieces) through the impact ions of Ar gas (plasma). The copper pieces are fixed on a different sample holder for each kind of component, inside of a class 100 cleanroom in order to handle the copper pieces in a controlled environment. The holder is placed in the vacuum chamber and before the plasma cleaning, 12~hours of baking at 100$^{\degree}$C is performed in order to degas adsorbed chemical substances and humidity. After the baking, a uniform magnetic field of 1.5x10$^{-2}$~T is generated and a 30~W power plasma is supplied for 5~minutes.

\end{itemize}

\section{Detector assembly}
\begin{figure}[h]
\centering
\includegraphics[width=1\columnwidth]{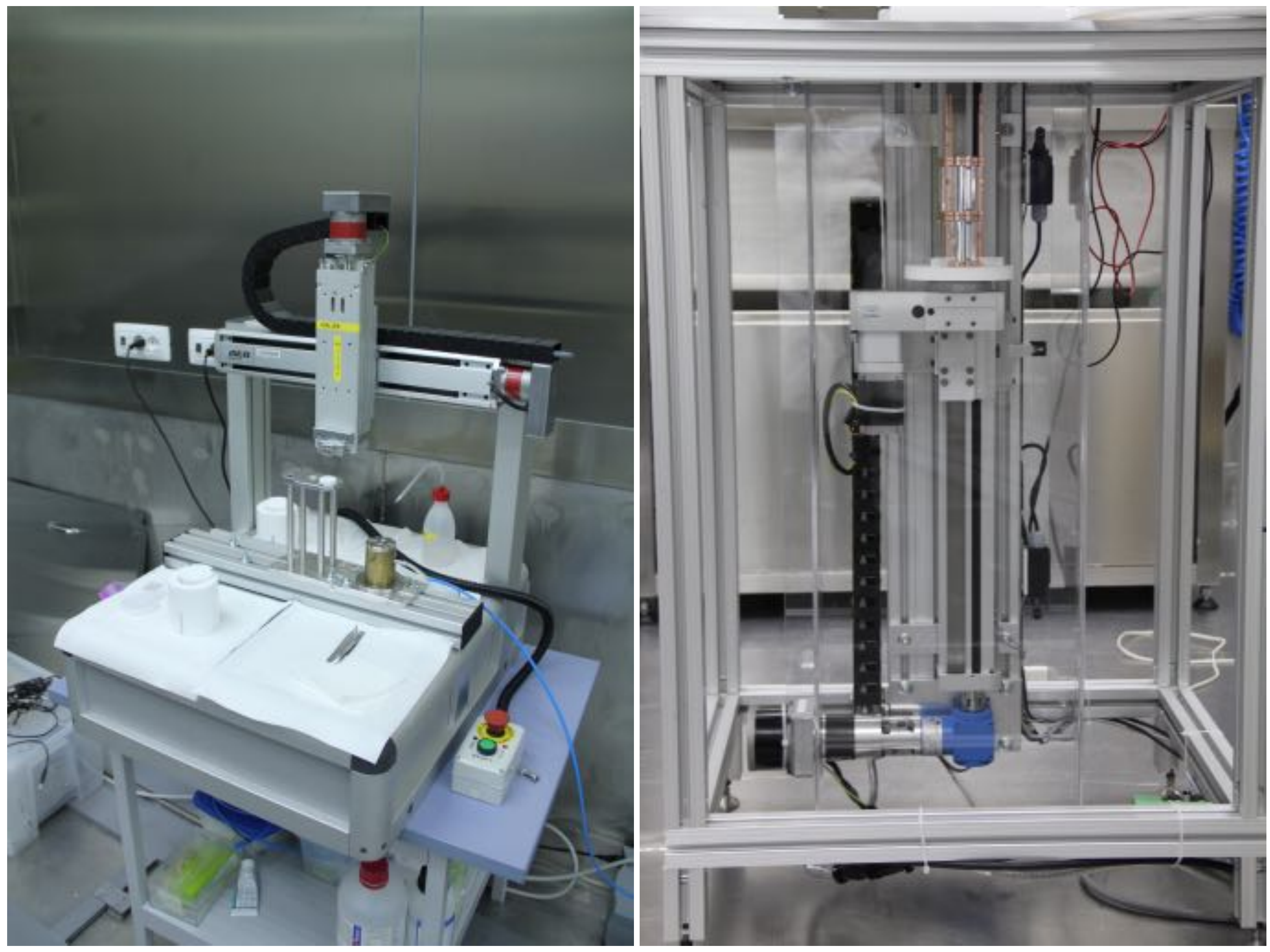}
\caption{\label{fig:cleanroom}Photos of the two working areas inside the cleanroom. On the left is shown the station were the Ge-NTD thermal sensors are glued to the absorbers, on the right is shown the area where the tower are assembled.}
\end{figure}

All activities for the construction of CUPID-0 detector were carried out in an underground Rn-suppressed clean room, with a Rn contamination of less than 5 mBq/m$^3$ located in the Hall C of the LNGS. The cleanroom contained two separate workstations: one for gluing the sensors to the ZnSe crystals and to the Ge wafers (see Fig.~\ref{fig:cleanroom}, left), and one for building and instrumenting the towers (see Fig.~\ref{fig:cleanroom}, right). A nitrogen-flushed storage container was also installed in the clean room for hosting the detector after its assembly.

\begin{figure}
\begin{centering}
\includegraphics[width=1\columnwidth]{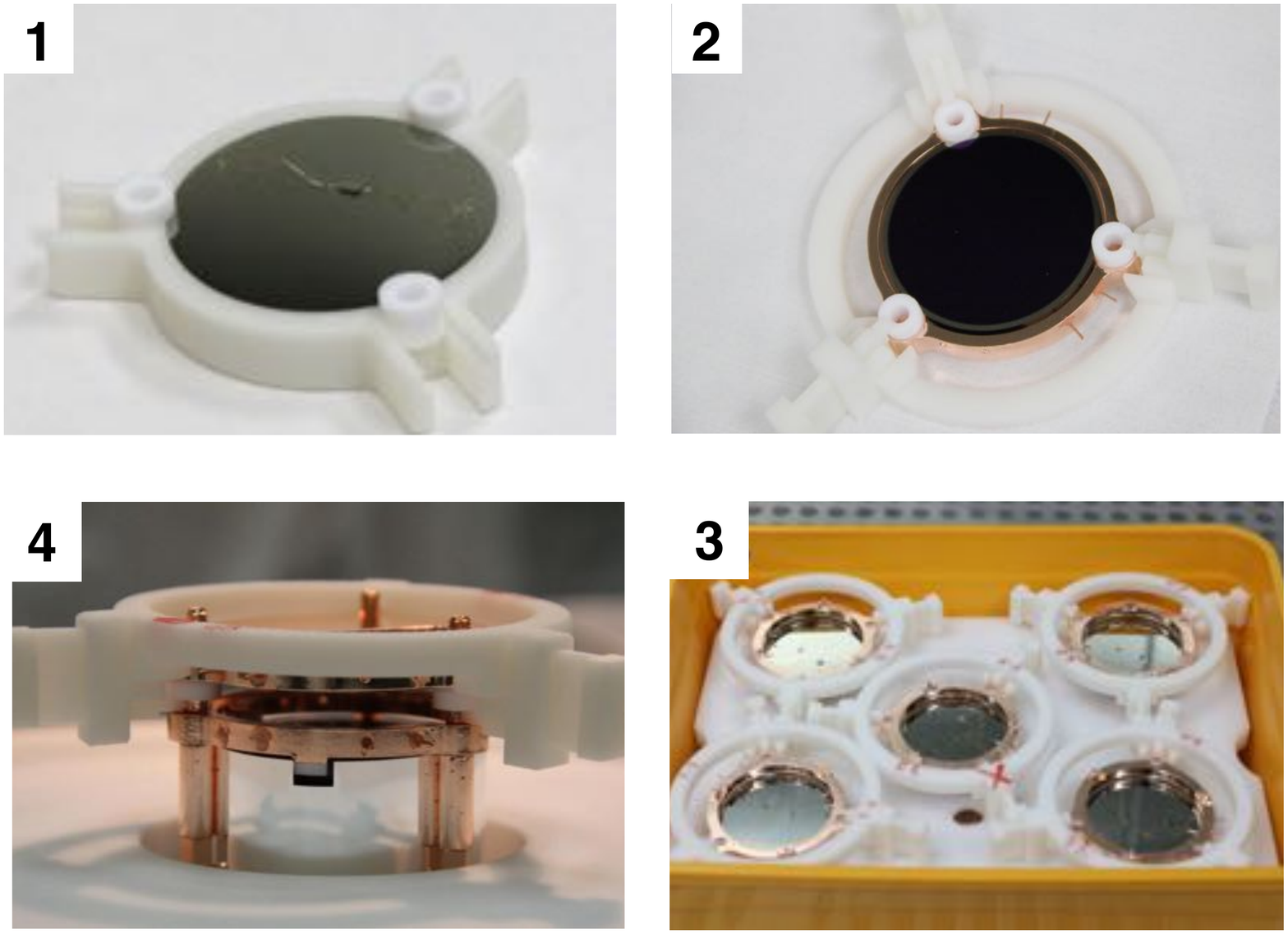}
\caption{Light detectors assembly: 1) Positioning of the Ge wafer with the PTFE O-shape holders on the mounting template. 2) Positioning of the copper holder and connection of the Ge-NTD gold wires for the detector read-out. 3) Storage of the assembled light detectors in a vacuum box. 4) Fit and tighten of the handling tool and removal from the mounting template.}
\label{fig:LD_assembly}
\end{centering}
\end{figure}

The first step in the CUPID-0 tower construction was the pre-assembling of the light detectors: each Ge wafer equipped with Ge-NTD was hold in its copper frame using three PTFE O-shape holders. Since the Ge wafers are 170~$\mu$m thick specific assembly tools were developed in order to avoid accidental damage and recontamination of the copper frames during the assembly procedures. These tools were made with a ENVISIONTEC ULTRA 3SP 3D printer using a highly radiopure plastic resin (see Tab.~\ref{tab:3D_radiopurity}). These tools consisted of a mounting template and of a handling system (see Fig.~\ref{fig:LD_assembly}-1 and ~\ref{fig:LD_assembly}-4 respectively). The assembly procedure for the 31 LDs of the CUPID-0 detector consisted in the following steps: positioning of the Ge wafer with the PTFE O-shape holders on the mounting 3D-printed template, installation of the Cu frame on the LD and connection of the Ge-NTD gold wires to the Cu frame for the sensor read-out, fitting and tightening of the handling tool and removal from the mounting template and finally their storage of the assembled light detectors in a vacuum box, see Fig.~\ref{fig:LD_assembly}.

\begin{table}
\begin{center}
\caption{Radioactive contamination of the polymer-resin used for the 3D-printing of the detector assembling tools (3SP WHITE D7). } 
\begin{tabular}{lccc}
\hline\noalign{\smallskip}
Chain & Nuclide  & Activity\\ 
            & & [mBq/kg] \\
\noalign{\smallskip}\hline\noalign{\smallskip}
$^{232}$Th & &  \\
 & $^{228}$Ra & $<$~9.3  \\
 & $^{228}$Th & $<$~10.3 \\ 
\noalign{\smallskip}\hline\noalign{\smallskip}
$^{238}$U & &  \\
 & $^{226}$Ra &  $<$~3.8 \\
 & $^{234}$Th & $<$~73 \\
 & $^{234m}$Pa & $<$~0.25 \\
\noalign{\smallskip}\hline\noalign{\smallskip}
$^{235}$U  & $^{235}$U &  $<$~5.3 \\
\noalign{\smallskip}\hline\noalign{\smallskip}
 & $^{40}$K & 81$\pm$35 \\
\noalign{\smallskip}\hline\noalign{\smallskip}
\end{tabular}
\label{tab:3D_radiopurity}
\end{center}
\end{table}

The second step in assembling the CUPID-0 detector was to physically assembly the towers using Cu, PTFE, 26 crystals and 31 pre-assembled LDs. The towers were manually built, one floor at the time~\footnote{A floor is defined as a single module, this is composed by a Zn$^{82}$Se and its most adjacent LD.}, starting from the lower one. In order to maintain a suitable operational working height, the towers were assembled in an automatically adjustable table, named garage. After the assembly of the first floor, the detector is lowered by the height of this floor, in this configuration the operator always works at the same level.

\begin{figure}
\begin{centering}
\includegraphics[width=1\columnwidth]{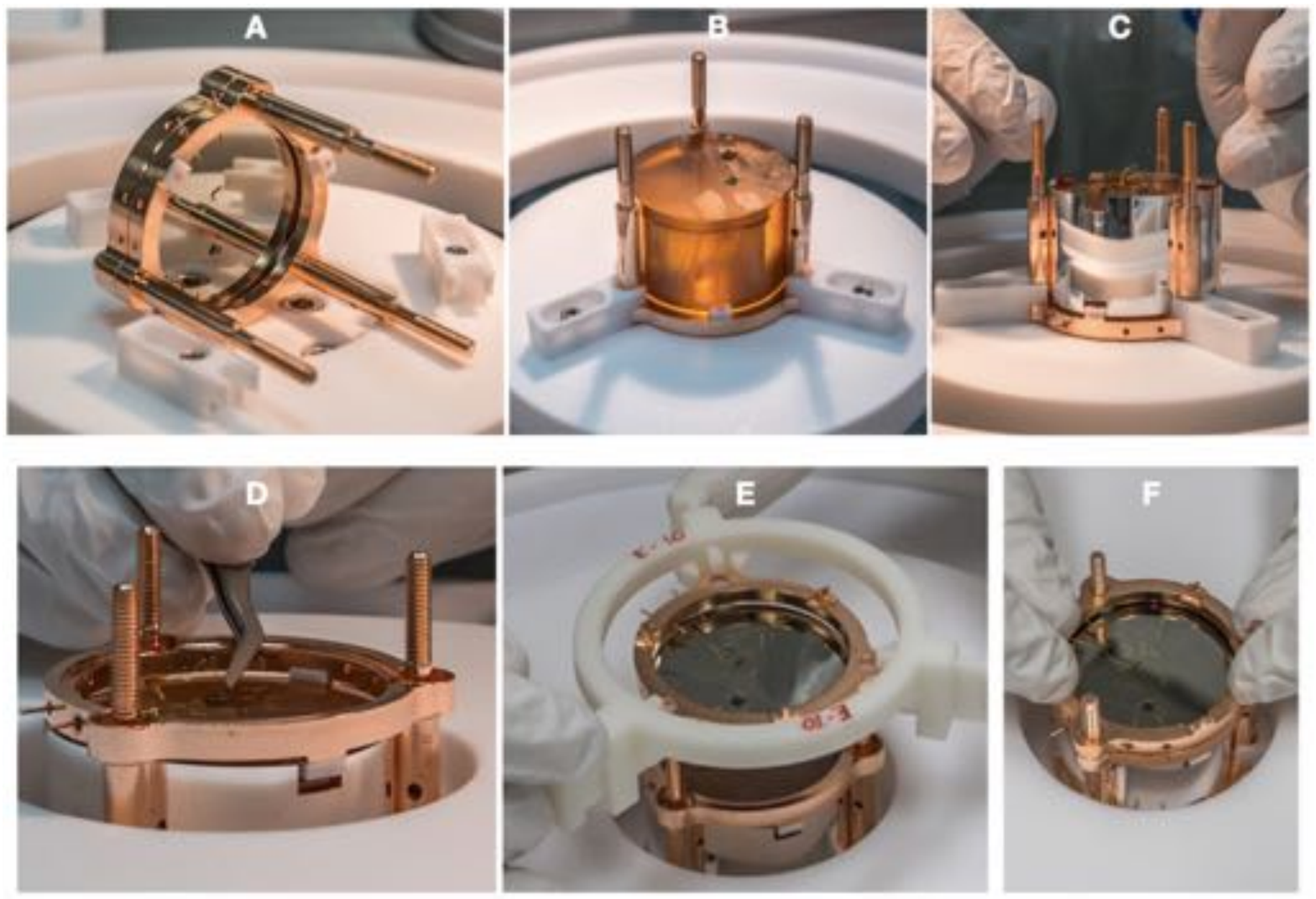}
\caption{Single tower assembly: A) Positioning of the first pre-assembled LD on the bottom copper holder of the tower; installation of the first three columns that will host the ZnSe crystal. B) Positioning of the ZnSe crystal on the bottom copper frame equipped with three S-shaped PTFE clamps. C) Positioning of the VIKUITI-3M reflective foil. D) Positioning of the top copper holder equipped with three S-shaped PTFE clamps; connection of the Ge-NTD gold wires in the inner copper pins. E) Positioning of the top pre-assembled LD. F) Coupling of the pre-assembled LD with the top ZnSe copper holder.}
\label{fig:tower_assembly}
\end{centering}
\end{figure}

The main steps for the assembly of a single tower are shown in Fig.~\ref{fig:tower_assembly}. The first operation is the positioning of the first pre-assembled LD on the bottom copper holder of the tower and the installation of the first three columns that will host the ZnSe crystal, then the positioning of the ZnSe crystal on the bottom copper frame equipped with three S-shaped PTFE clamps. The third step is the installation of the VIKUITI-3M reflector and the top copper holder equipped with three S-shaped PTFE clamps. At this point the electrical connections between the gold-wire and the mechanical structure are finalized. The wires are crimped into the insulated copper tubes glued into the frames. Finally the second LD is installed on the top frame of the previously installed ZnSe. All these procedure are repeated until a tower of 5 floors is completed.

\begin{figure}
\begin{centering}
\includegraphics[width=1\columnwidth]{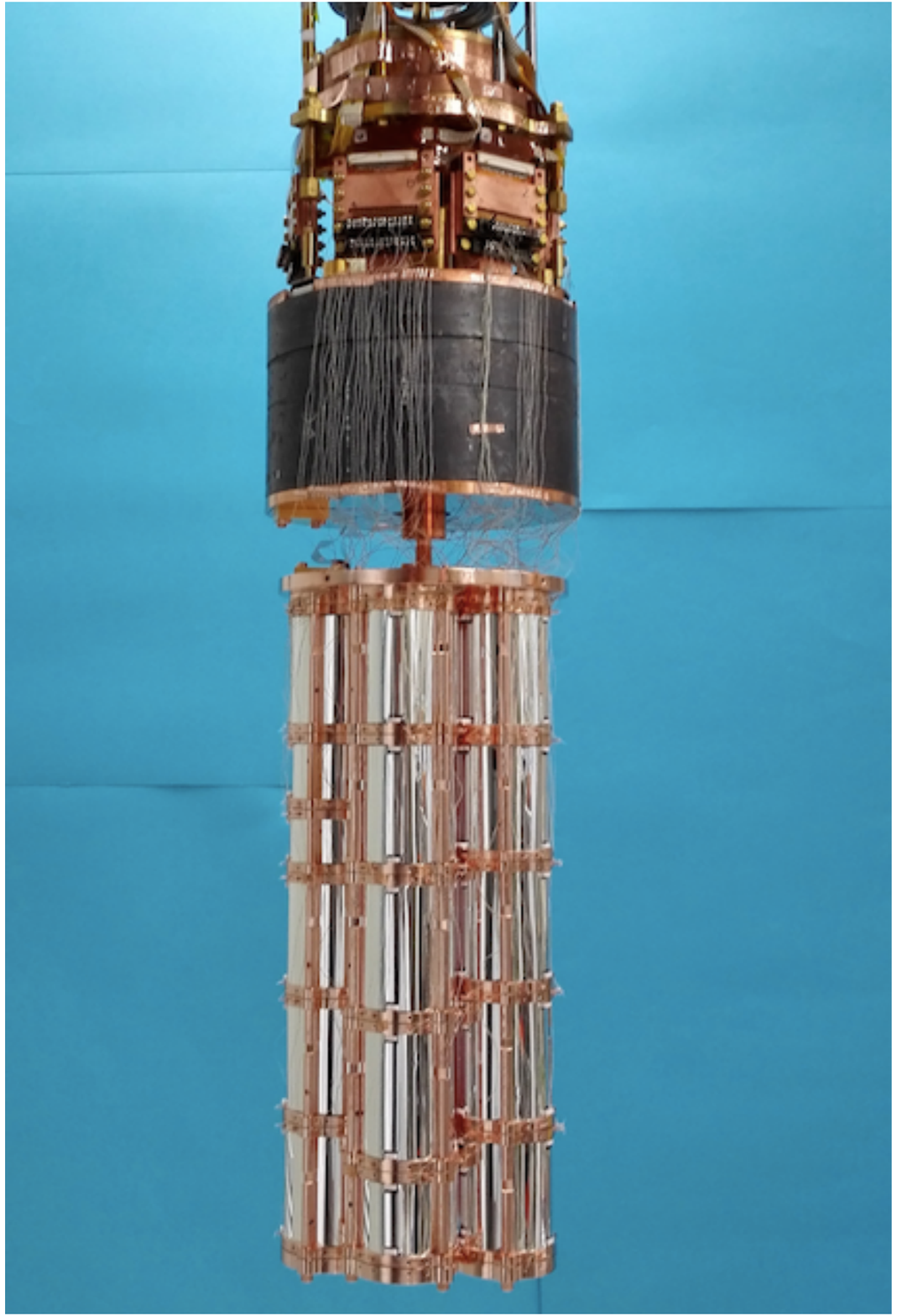}
\caption{CUPID-0 detector installed on the cryogenic system, just below the 10~cm thick Roman Pb shielding.}
\label{fig:final_detector}
\end{centering}
\end{figure}

After the completion of the detector assembly the five towers are hosted on a copper plate which acts as support structure and connection between the cryogenic system and the detectors, see Fig.~\ref{fig:final_detector}. 

\section{Cryostat}

CUPID-0 cryostat is the same cryogenic infrastructure that hosted the CUORICINO~\cite{Qino} and the CUORE-0~\cite{CUORE-0} detectors. This system was upgraded in order to meet our stringent requirements in terms of low vibrational environment and increase number of read-out channels. 
The cryostat was commissioned at the LNGS in 1988 and it is a Oxford TL1000 with a copper He dewar. The $^{3}$He/$^{4}$He dilution unit has a cooling power at 100~mK of about 1~mW, this ensures the possibility to install in the system a large number of read-out channels without spoiling the cryostat performance in terms of cooling power.

\begin{figure}
\begin{centering}
\includegraphics[width=1\columnwidth]{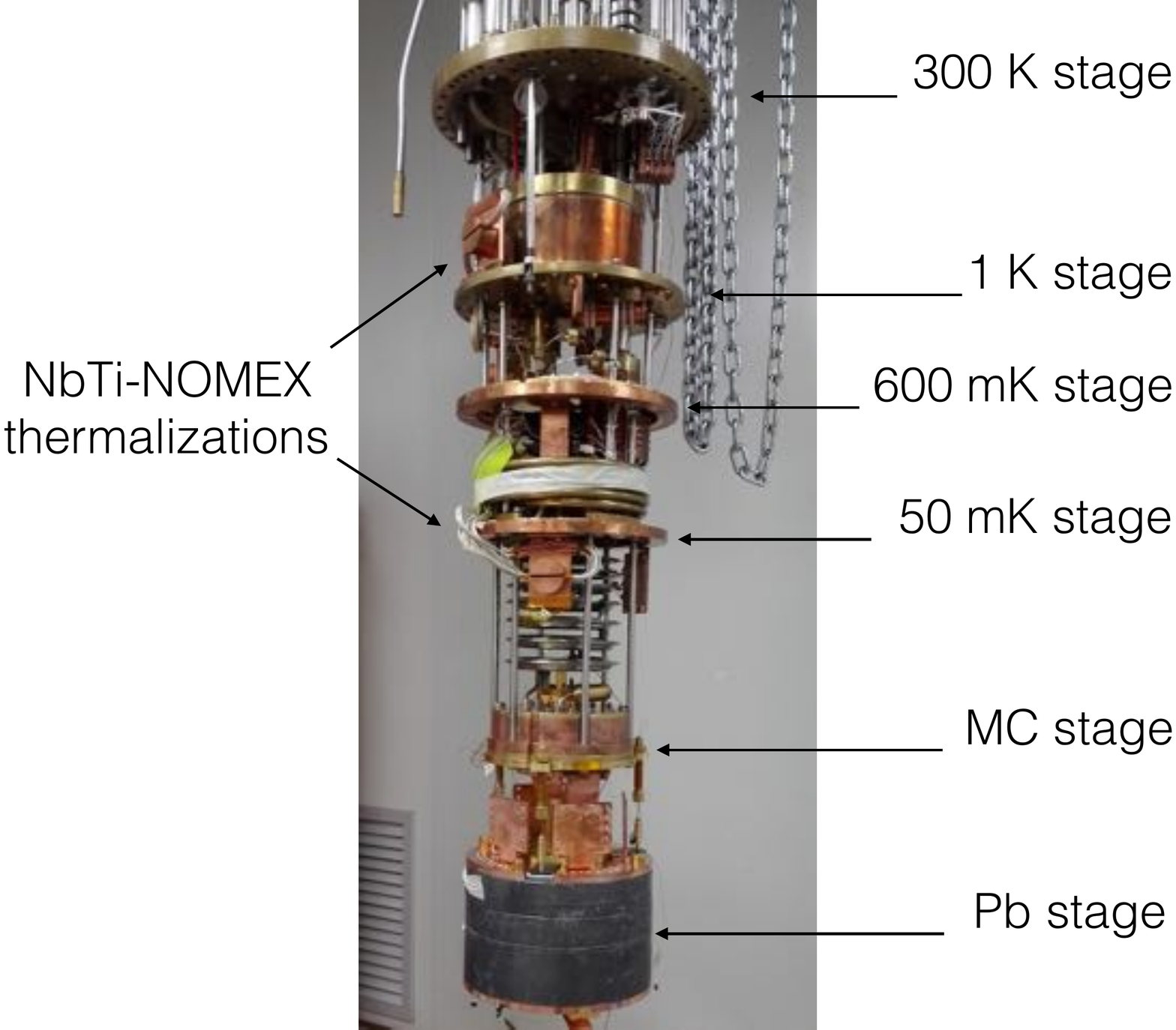}
\caption{Dilution unit photo. The different thermalization stages are identified by the arrows. A Roman Pb shield is hanged on the MC stages by means of a vibrational damping system. On the Pb stage are installed the junction board shown in Fig.~\ref{fig:zif}.}
\label{fig:cryo}
\end{centering}
\end{figure}

In Fig.~\ref{fig:cryo}, the cryogenic system hosting the CUPID-0 detector is shown. The detector is installed right below a Roman Pb shield by means of a spring and it is thermally coupled to the Mixing Chamber stage (MC) by means of a high-purity (99.999\%) copper foil of 50~$\mu$m thickness. The MC, which is the coldest point of the system, ensures the detector cooling down to the designed base temperature of 7.5~mK.

Two major upgrades were implemented in the system compared to the previous configuration: increasing the number of read-out channels and installing a mechanical decoupling system as anti-vibrational damping system. Upgrading the number of read-out channels was a mandatory step given the fact that for each ZnSe crystal there is also a LD, this doubles the number of wires from room temperature down to the detectors. When hosting the CUORE-0 experiment, the cryostat was able to read out up to 52 bolometers, now the system can handle up to 136 detectors, irrespectively if they are ZnSe crystals or LDs. For CUPID-0 67 channels are used: 26 for ZnSe crystals, 31 for LDs and 10 thermometers for monitoring the stability of the detectors and of the system. The remaining available channels might be employed for a future upgrade of the detector. 

\begin{figure}
\begin{centering}
\includegraphics[width=1\columnwidth]{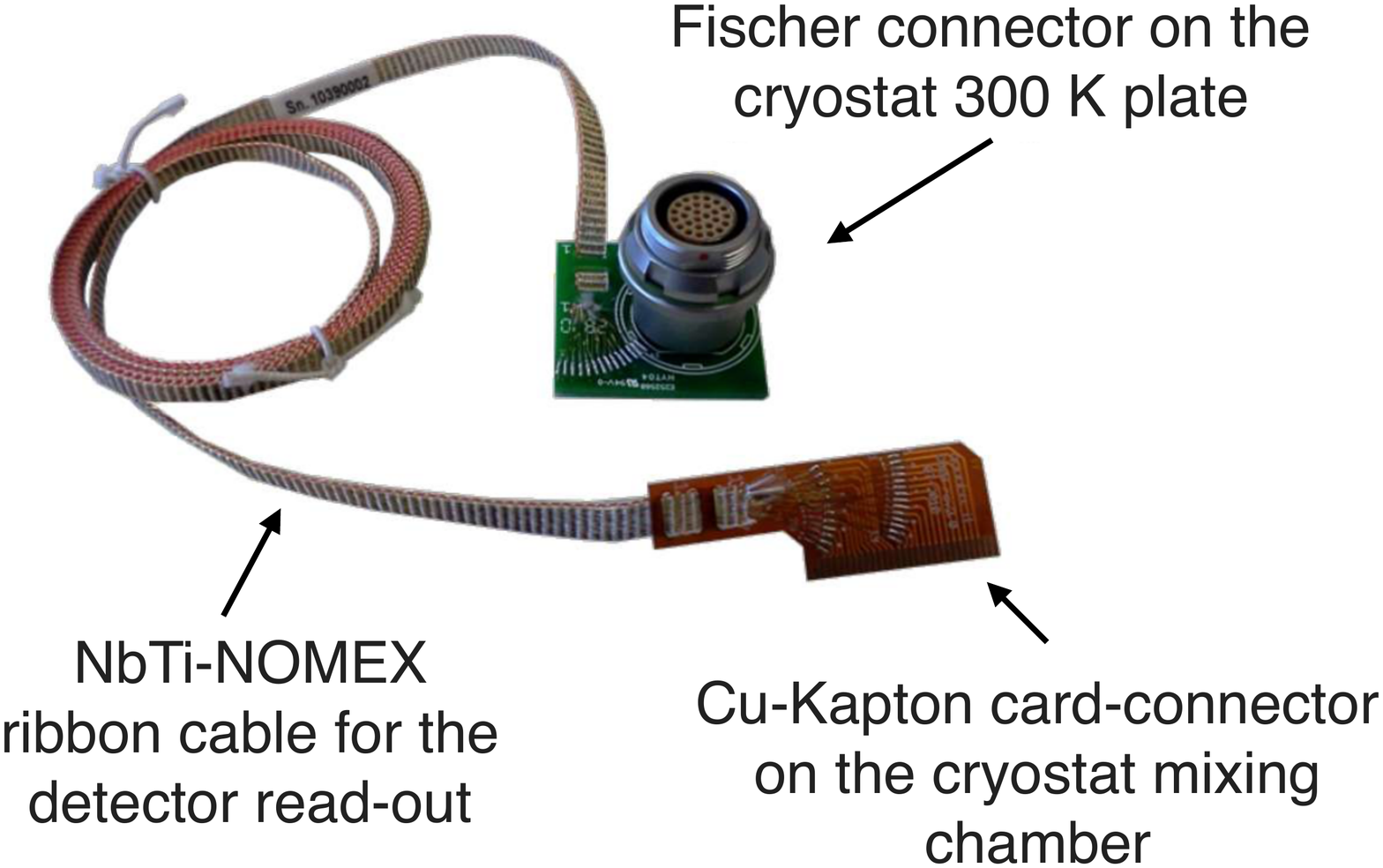}
\caption{NbTi-NOMEX cable from 300~K to the mixing chamber. On the room temperature side they are soldered to Fischer 27-pin connectors on the other hand they are soldered to customized Cu-Kapton Zero Insertion Force connectors.}
\label{fig:nomex}
\end{centering}
\end{figure}

\begin{figure}
\begin{centering}
\includegraphics[width=1\columnwidth]{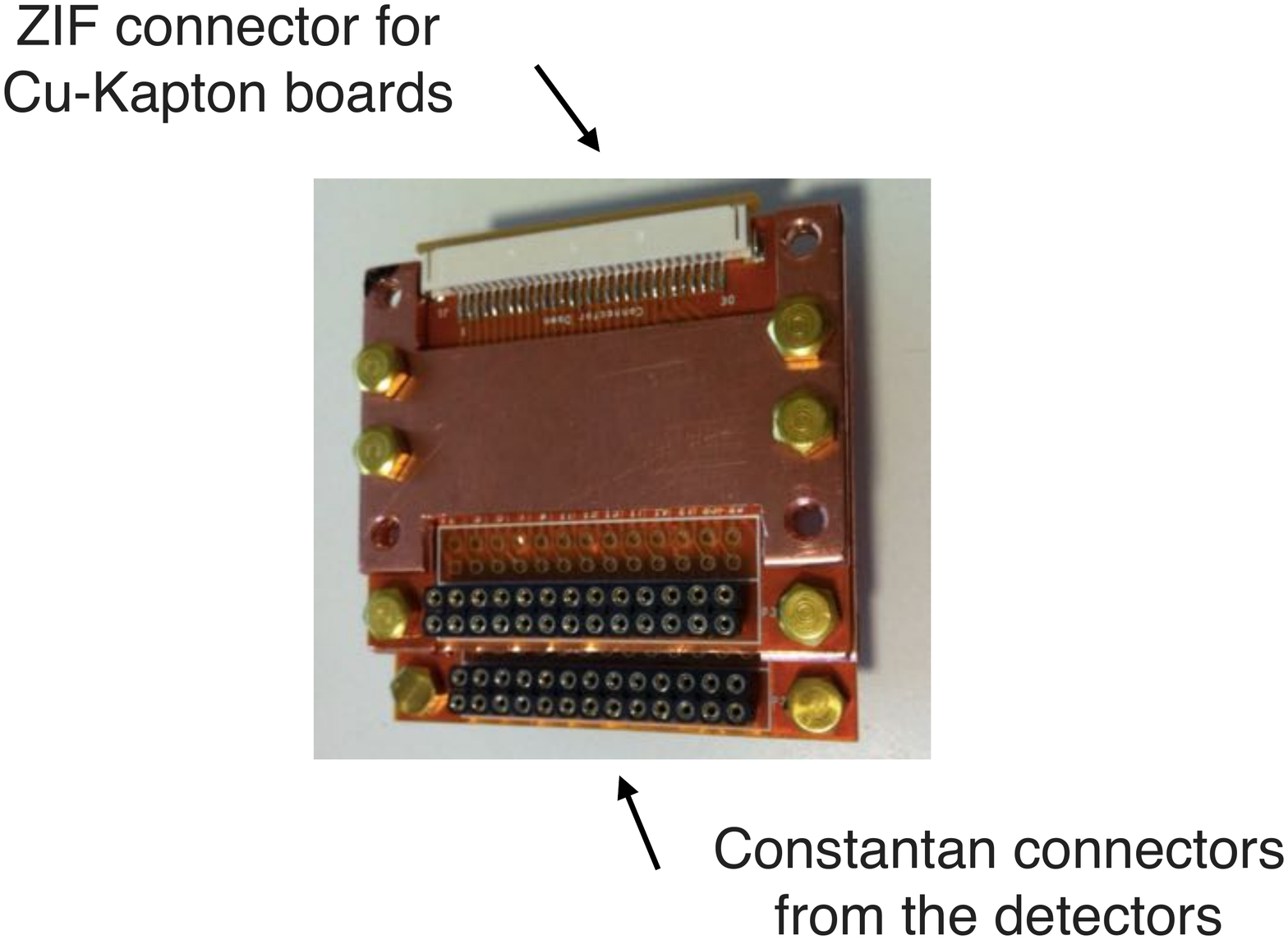}
\caption{Junction boards on the MC for connecting the NbTi-NOMEX cables to the constantan wires from the detectors.}
\label{fig:zif}
\end{centering}
\end{figure}

The signals are extracted from the detectors using NbTi-NOMEX$^{\textregistered}$ ribbon cables. The NbTi-NOMEX cables run from room temperature down to the MC, see Fig.~\ref{fig:nomex}, while from the MC to the detectors there are twisted 60~$\mu$m constantan wires. All the cables are thermalized at the different temperature stages of the cryostat, and on the MC they are plugged into custom-made junction boards through Zero Insertion Force (ZIF) connectors, which connect the ribbon to the constantan wires, as shown in Fig.~\ref{fig:nomex} and Fig.~\ref{fig:zif}.

The NbTi-NOMEX ribbon cables are made of 13 twisted pairs of 100~$\mu$m NbTi twisted wires. Their low radioactivity and electrical properties~\cite{ribbon} make them the best choice for the detector read-out. They are characterized by low thermal conductivity, becoming superconducting below 10~K, low parasitic capacitance (100~pF/m) and a negligible cross-talk level, 500~twisting/m.

The second major upgrade of the cryogenic system consisted in a mechanical double stage anti-vibrational decoupling system, similar to the one developed in~\cite{salaC}. The main purpose for the development and installation of such system was driven by the fact that any microphonic noise source has to be minimized, in order to prevent any spoiling of the LD bolometric performance. In fact the LDs, having a higher sensitivity compared to the ZnSe crystals, requires a much lower vibrational environment compared to a system where only massive crystals are operated~\cite{CUORE_cryo}.

\begin{figure}
\begin{centering}
\includegraphics[width=1\columnwidth]{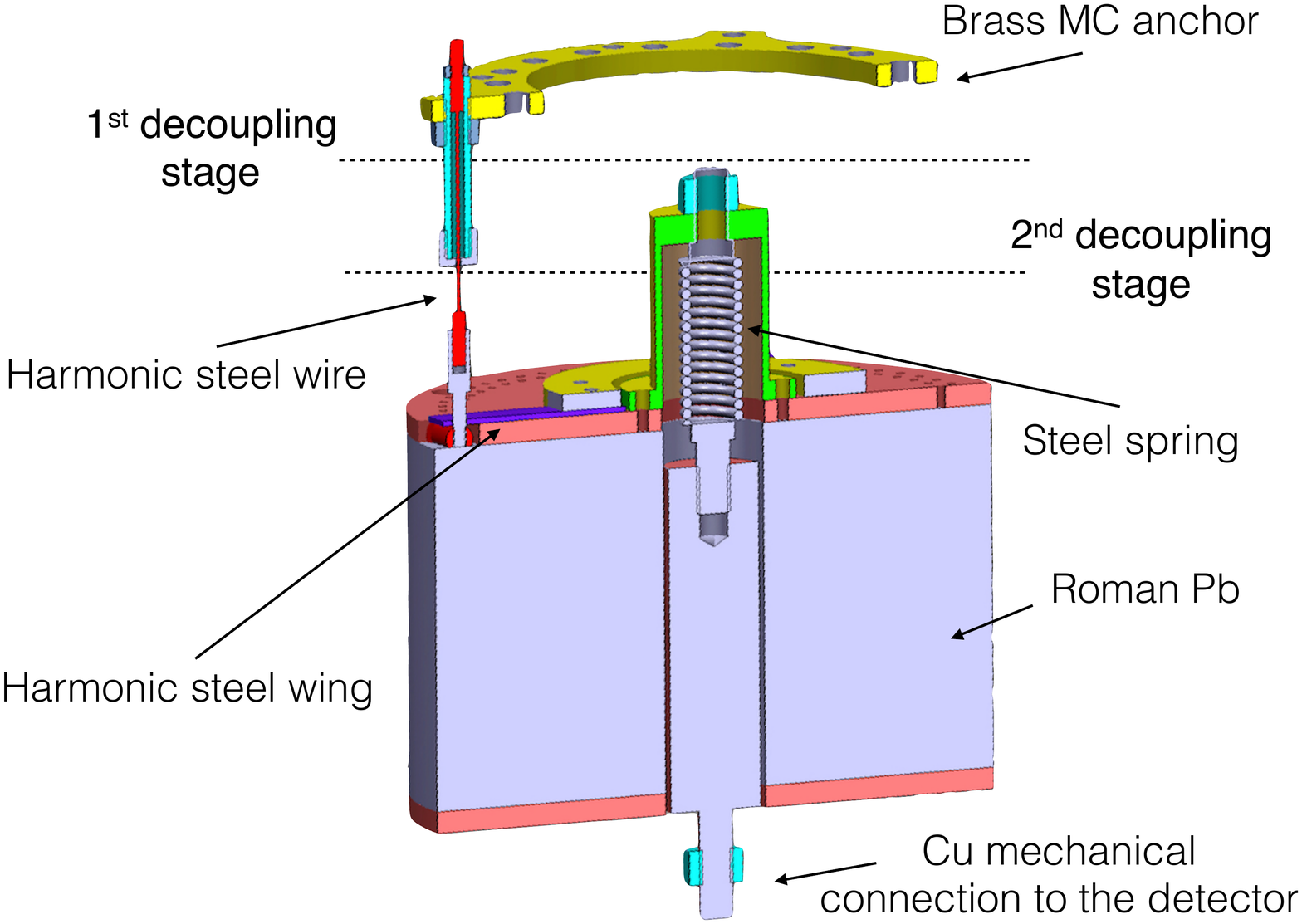}
\caption{Rendering of the double stage mechanical decoupling system. This is installed directly on the cryostat mixing chamber with the top brass anchor and on the bottom part there is a Cu mechanical connection for the detector installation.}
\label{fig:decoupling}
\end{centering}
\end{figure}

Fig.~\ref{fig:decoupling} shows a scheme of the mechanical decoupler which is directly hanging from the MC using the top brass ring. The circular brass piece holds in place the 10~cm thick Roman Pb shield by means of three custom-designed wires made of harmonic steel (red color of Fig.~\ref{fig:decoupling}, where just one wire connector is shown). The Roman Pb shield is connected to the harmonic steel wires by means of three harmonic steel wings mechanical anchored on the Pb (purple color of  Fig.~\ref{fig:decoupling}). The characteristic longitudinal resonance frequency of this first decoupling stage is about 12~Hz. On the top of the Pb shield is encapsulated a steel spring which is mechanical connected to the detectors by means of Cu cylinder housed inside the Pb shield. The Cu connector is mechanical decoupled from the Pb acting as second mechanical decoupling system, and it is characterized by longitudinal resonance frequency of about 5~Hz. The Cu connector acts as shielding from the radioactivity of the spring, which due to mechanical reasons is not made from high purity materials.

The thermal connection is ensured by several 100~$\mu$m thick (99.999 \% purity) Cu stripes between the MC and the first damping stage and by 2 -softer- 50~$\mu$m thick 6~cm long, 2~cm wide copper stripes between the first damping stage and the Cu detector top plate.

\section{Detector readout}\label{sec:readout}

\begin{figure}[h]
\centering
\includegraphics[width=1\columnwidth]{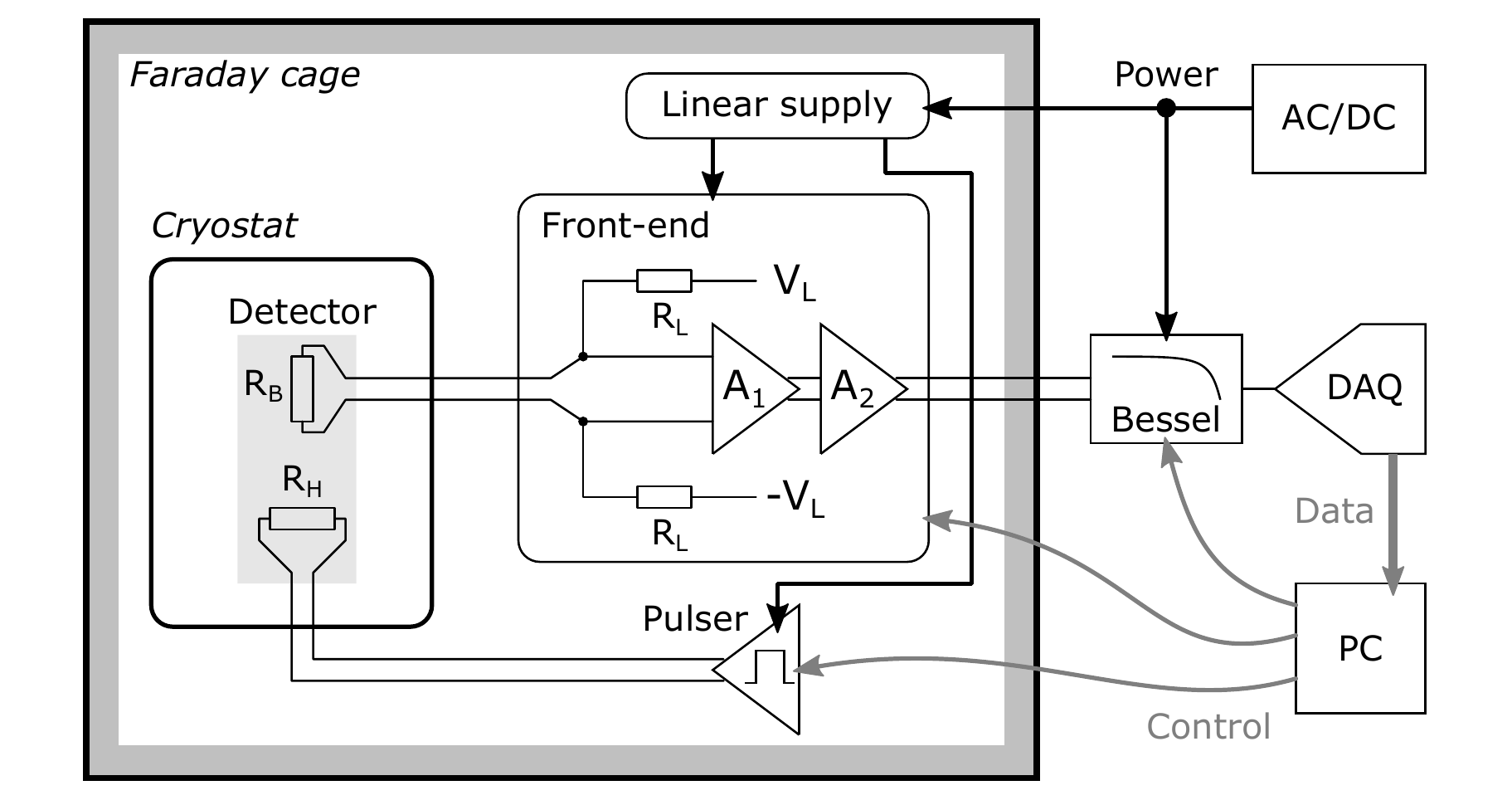}
\caption{\label{fig:readout}Block diagram of a readout channel, from the detector to the DAQ.}
\end{figure}

The readout system of CUPID-0 shares the same general structure as that of the CUORE experiment \cite{electronics:frontend}.
Many of the operating parameters were optimized for CUPID-0.
Fig.~\ref{fig:readout} shows the block diagram of the readout chain for a single detector, valid for both crystals and LDs.
The thermistor $R_{B}$ is biased with a DC current through a pair of load resistors $R_L= 30\ G\Omega$ (10~G$\Omega$).
The bias generator $V_L$ can be set between $-25\ V$ and $25\ V$ with 16-bit resolution.
The voltage across the thermistor is amplified by the two-stage amplifier $A_1$ and $A_2$.
The total gain can be set between $27\ V/V$ and $10000\ V/V$ with 12-bit resolution.
The input stage of $A_1$ is based on a JFET differential pair.
We designed two different options:
a) low parallel noise, less than $100\ fA$ below $50\ ^{\circ}C$, and about $3.5\ nV/\sqrt{Hz}$ series white noise (twice this value at $1$~Hz);
and b) low series noise, $1.2\ nV/\sqrt{Hz}$ (twice this value at $1\ Hz$), with larger parallel noise.
In principle, the choice depends on the value of detector impedance.
In practice we observed that in both cases this stage was not limiting the resolution, and option a) was found adequate for all detectors.
The thermal parallel noise of the load resistors $R_L$, whose maximum value and temperature of operation are constrained by practicality.

\begin{figure}[b]
  \begin{subfigure}[b]{0.49\columnwidth}
    \includegraphics[width=\columnwidth]{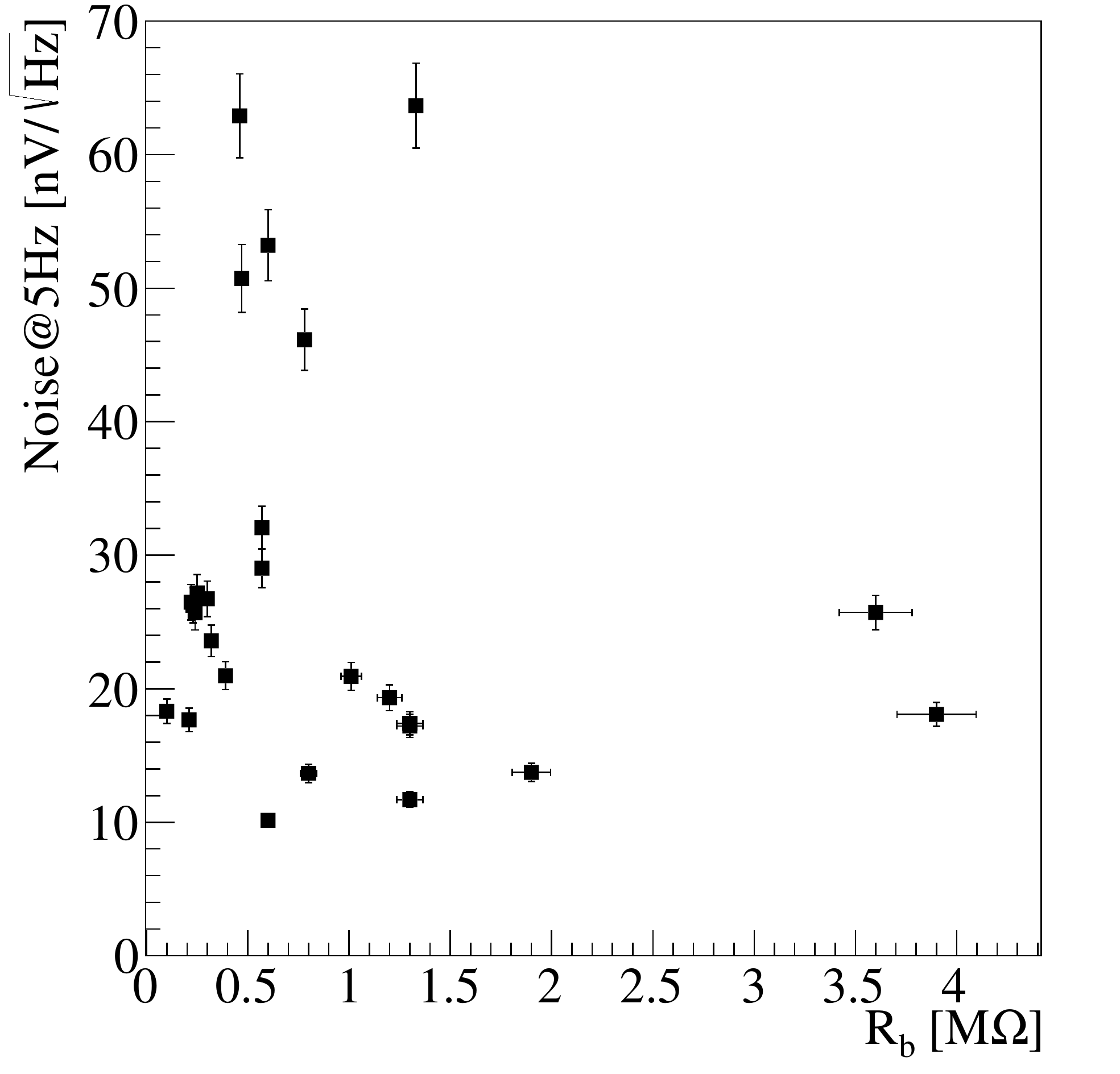}
    \caption{RMS noise at 5~Hz for the ZnSe detectors as a function of the dynamic bolometric impedance of the Ge-NTD.}
    \label{fig:Lorenzo1}
  \end{subfigure}
  \hfill
  \begin{subfigure}[b]{0.49\columnwidth}
    \includegraphics[width=\columnwidth]{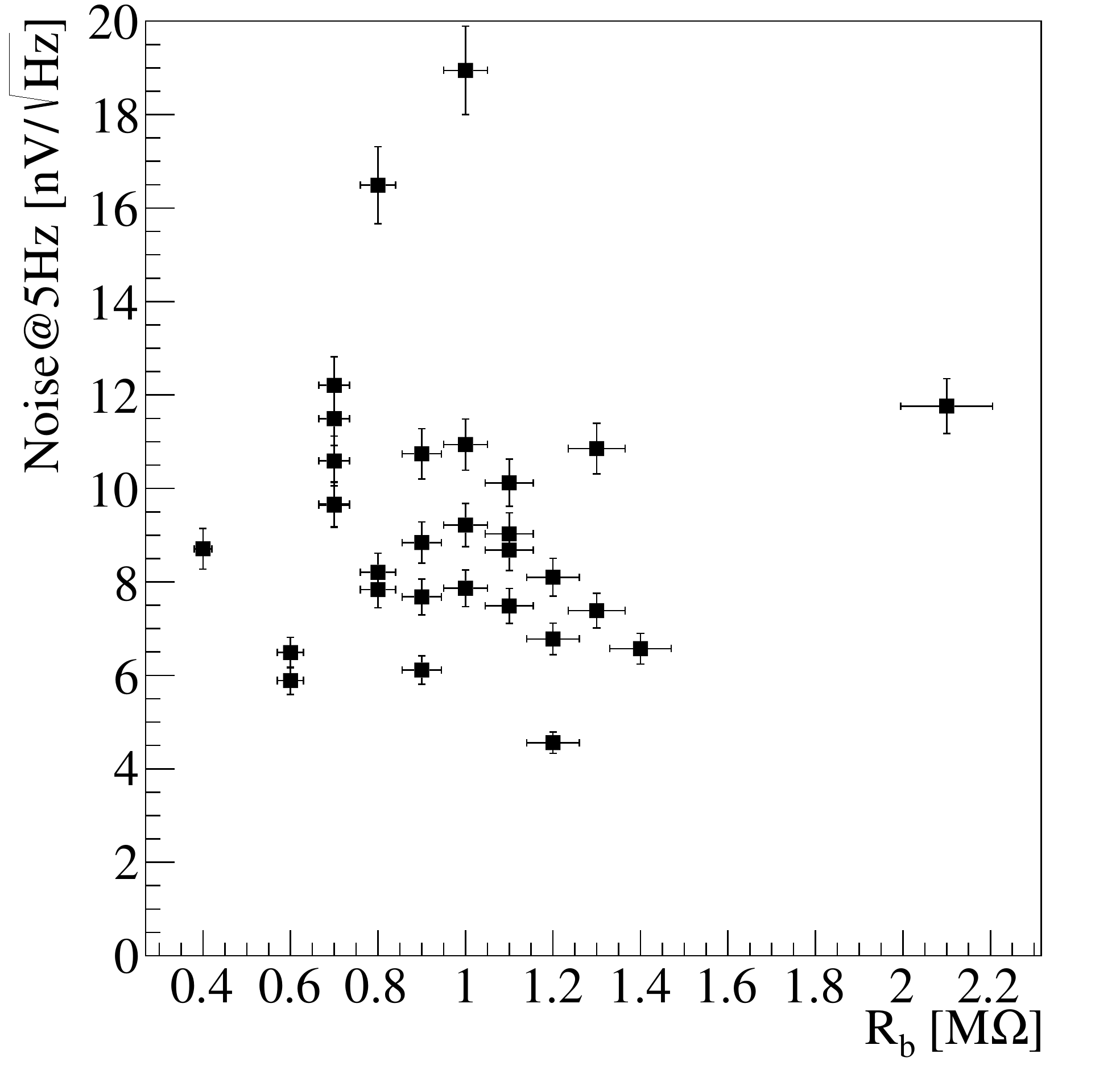}
    \caption{RMS noise at 5~Hz for the ZnSe detectors as a function of the dynamic bolometric impedance of the Ge-NTD}
    \label{fig:Lorenzo2}
  \end{subfigure}
  \caption{Detectors RMS noise at 5~Hz as a function of the Ge-NTD dynamic impedance.}
\end{figure}

In Fig.~\ref{fig:Lorenzo1} and Fig.~\ref{fig:Lorenzo2}, the RMS noise at 5 Hz is shown for the ZnSe crystals and the LD, respectively. The measured noise in the present setup is larger than expected considered the value of the detector impedances, the contributions from the front-end amplifier and the load resistors. Future optimization will be focused on further reducing the observed noise, attempting to reach this limit.\newline

The amplified signals are routed out of the Faraday cage to the antialiasing filters (Bessel-Thomson, 6 poles) and the data acquisition system (DAQ) \cite{electronics:bessel}.
The cutoff frequencies, settable in 4 steps, are $15, 35, 100, 120$~Hz for the crystals and $15, 100, 140, 220$~Hz for the LDs. 

The Pulser board is used to generate voltage pulses, which are injected onto the detector by resistor $R_H$ \cite{electronics:pulser}.
The pulses are triggered and tagged by the DAQ, and used for relative calibration during data taking.
Their noise is negligible, typically at the order of $10\ $ppm RMS, and their thermal stability is better than 1 ppm$/^{\circ}C$, reducing the need for calibration runs with radioactive sources.
Similar boards are also used to stabilize the temperature of the mixing chamber and of the detector holder through PI (proportional-integral) control loops.
The power supply is provided by a two stage system: a commercial floating AC/DC generator with a custom filtering solution  \cite{electronics:acdc}, followed by two custom linear power supplies with low noise ($1.6\ \mu V$ peak to peak between $0.1$~Hz and $100$~Hz) and high stability (about $1\ $ppm$/^{\circ}C$), which serve also as reference voltages for the front-end amplifier and the bias generator \cite{electronics:linear}.
In this way the entire system is able to maintain a stability better than $10\ $ppm$/^{\circ}C$.
The front-end and the Pulsers are housed in 19'' 6U and 19'' 3U standard racks respectively.
A total of 66 channels are available, which are also used to read out diagnostic thermometers.
Two Pulser boards (4 channels each) are used for stabilization, and one (two channels) is used for PI control.
The entire system is remotely controlled by a PC through an optically coupled CAN bus.

\section{Detector configuration}
The detector was cooled down in February 2017 and the first operation to be performed was the evaluation of the detector temperatures. This consists in measuring the resistance of the Ge-NTD sensors on all the detectors. According to the sensor design we are expecting a base resistance ($R_{base}$) of hundreds of M$\Omega$ on the ZnSe and one at least order of magnitude larger values for the LD, given the reduced sensor mass. The spread in the distribution of the base temperature provides information on the uniformity of the absorber properties, namely the heat capacity of the system: ZnSe + Ge-NTD. This discrepancy can be stressed if looking at the distribution of the working resistance ($R_{work}$) of the detectors, the value of the resistance (or temperature) once the operational condition of the detector are set. 

\begin{figure}[!tbp]
  \begin{subfigure}[b]{0.49\columnwidth}
    \includegraphics[width=\columnwidth]{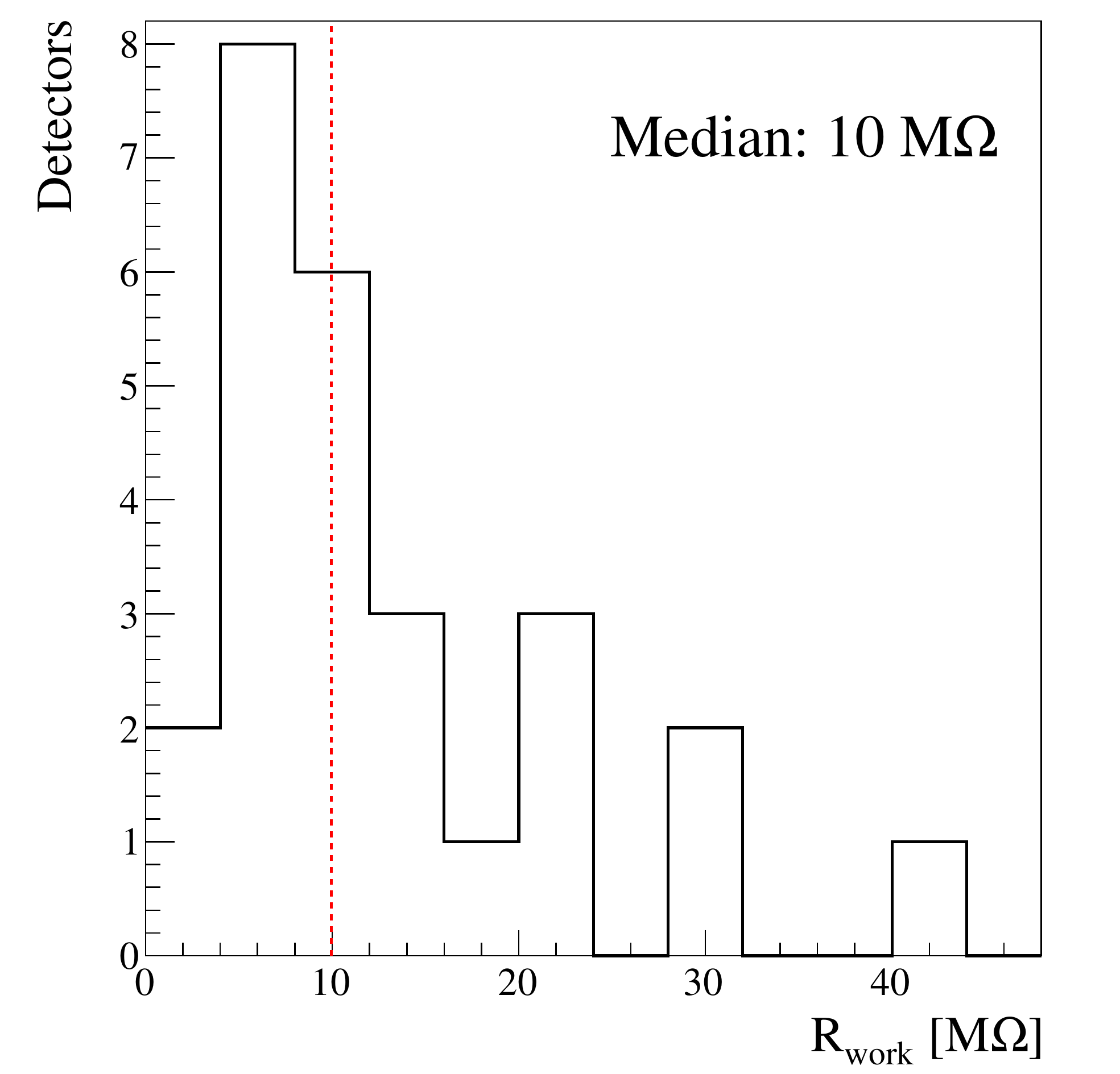}
    \caption{Ge-NTD resistance distributions for the ZnSe crystals.}
    \label{fig:Rworka}
  \end{subfigure}
  \hfill
  \begin{subfigure}[b]{0.49\columnwidth}
    \includegraphics[width=\columnwidth]{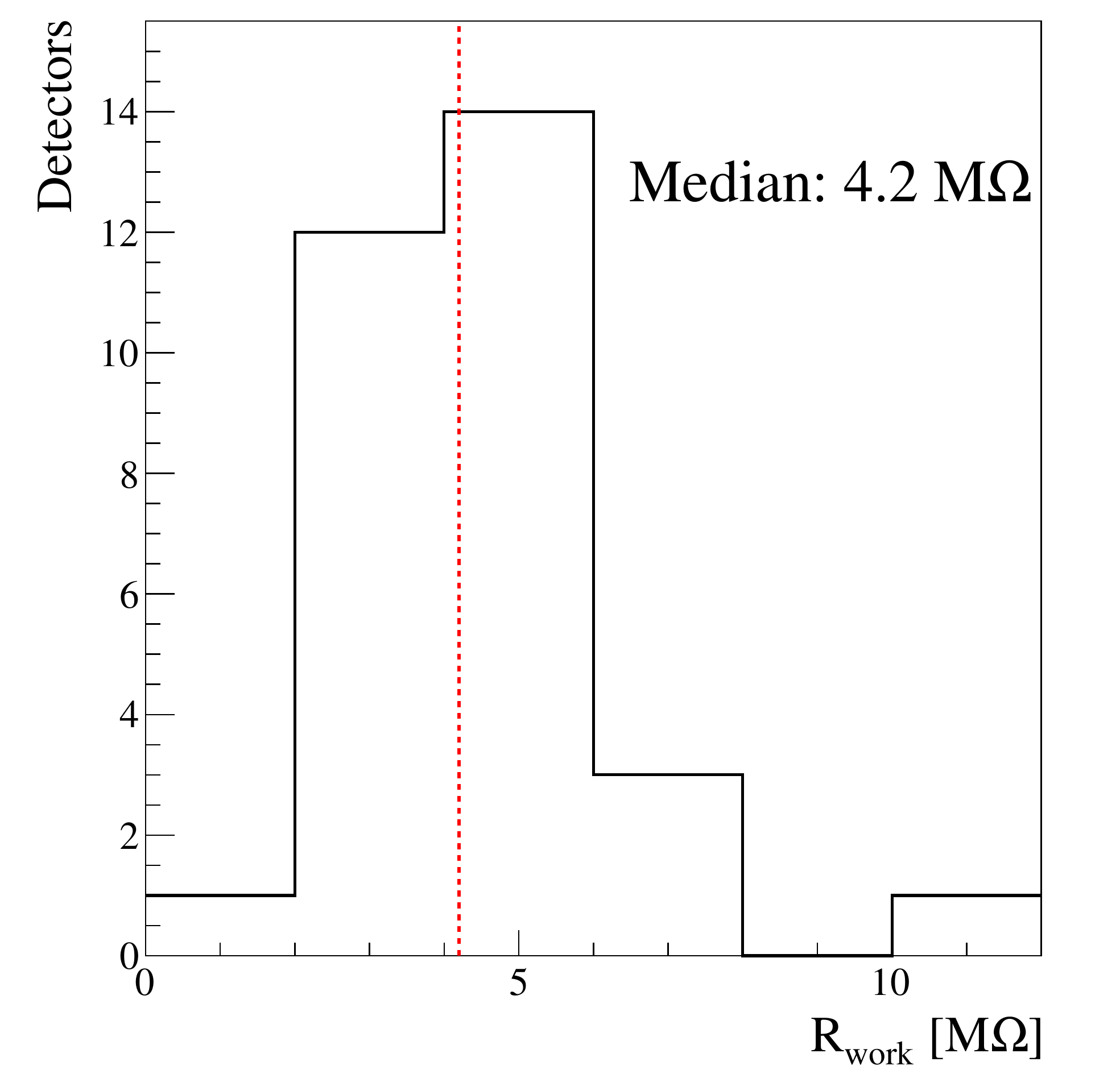}
    \caption{Ge-NTD resistance distributions for the LDs.}
    \label{fig:Rworkb}
  \end{subfigure}
  \caption{Distribution of the Ge-NTD sensor resistances for ZnSe (left) and LD (right) at the operating conditions.}
\end{figure}

\begin{figure}
\begin{centering}
\includegraphics[width=1.0\columnwidth]{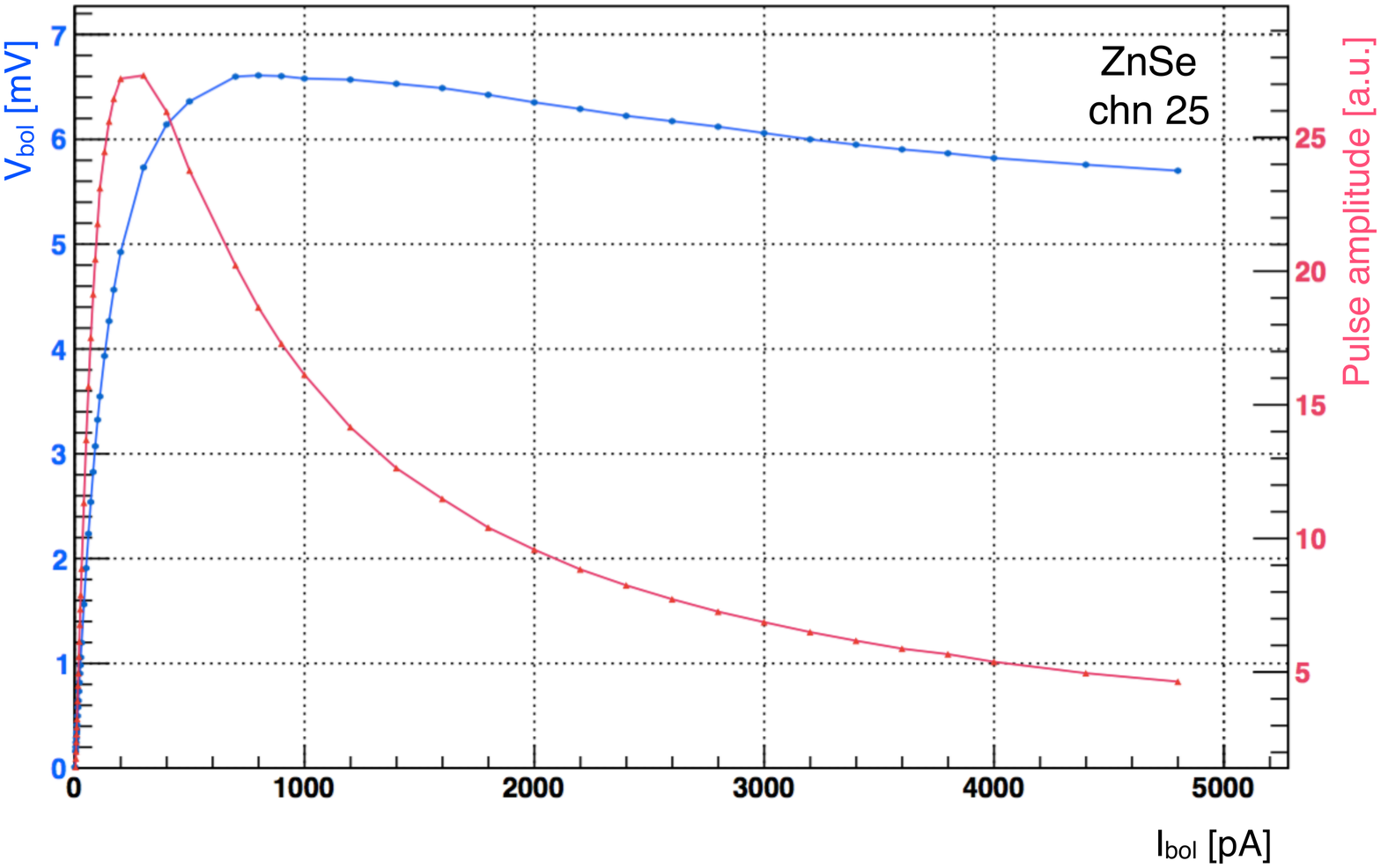}
\caption{Characteristic load curve of a CUPID-0 ZnSe crystal operated with a Ge-NTD thermal sensor. The figure shows how varying the detector biasing voltage acts on the signal amplitude (red) and on the voltage drop across the sensor resistance (blue).} 
\label{fig:loadcurve_vi}
\end{centering}
\end{figure}

\begin{figure}
\begin{centering}
\includegraphics[width=1.0\columnwidth]{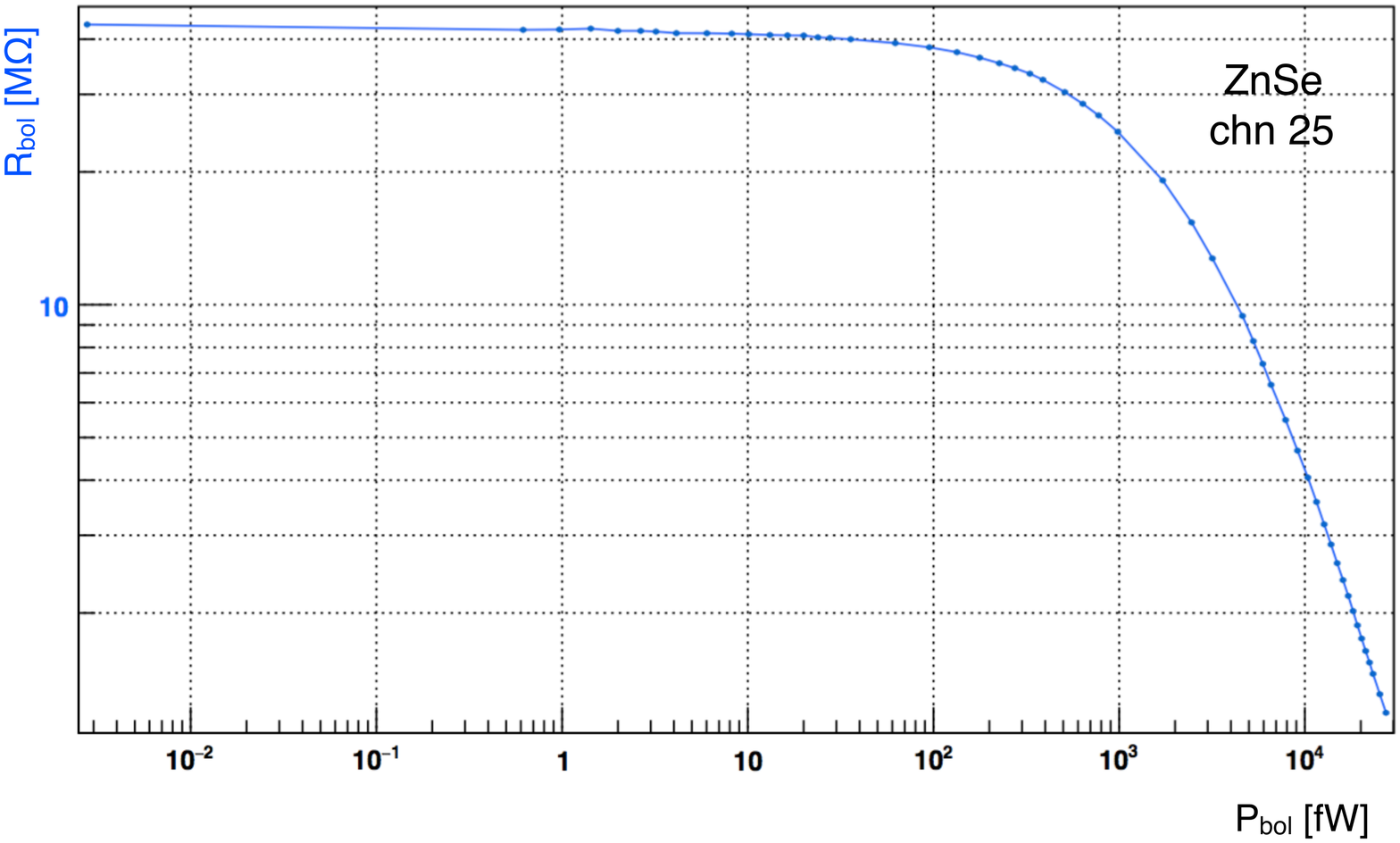}
\caption{Characteristic load curve of a CUPID-0 ZnSe crystal operated with a Ge-NTD thermal sensor. The figure shows how a Ge-NTD stands high power dissipation without affecting the sensor operational conditions.}
\label{fig:loadcurve_rp}
\end{centering}
\end{figure}

In Fig.~\ref{fig:Rworka} and Fig.~\ref{fig:Rworkb}, the distributions for the $R_{work}$ for the ZnSe and LDs are shown. The distribution of the LD resistances tells us that the production of such detectors is highly reproducible and there is a robust control of the critical aspects for the detector production. For the ZnSe, on the other hand the spread of the distribution is large and this is due to two different aspects: the first is because the crystals have different masses, hence different heat capacities, and the second because there is limited control of the ferromagnetic impurities inside the absorber. We performed a bias scan in order to evaluate the best operating condition of each detector, this was done varying the bias current of each detector and evaluating the best signal-to-noise ratio for each configuration. In Fig.~\ref{fig:loadcurve_vi}-\ref{fig:loadcurve_rp}, we show how the signal amplitude varies as a function of the detector biasing voltage for a ZnSe detector and how the sensor is able to stand sizeable power dissipation without affecting its working resistance. In the detectors the reference signals are generated by a Si resistor coupled to the crystal operated as Joule heater.

\begin{figure}
\begin{centering}
\includegraphics[width=1.0\columnwidth]{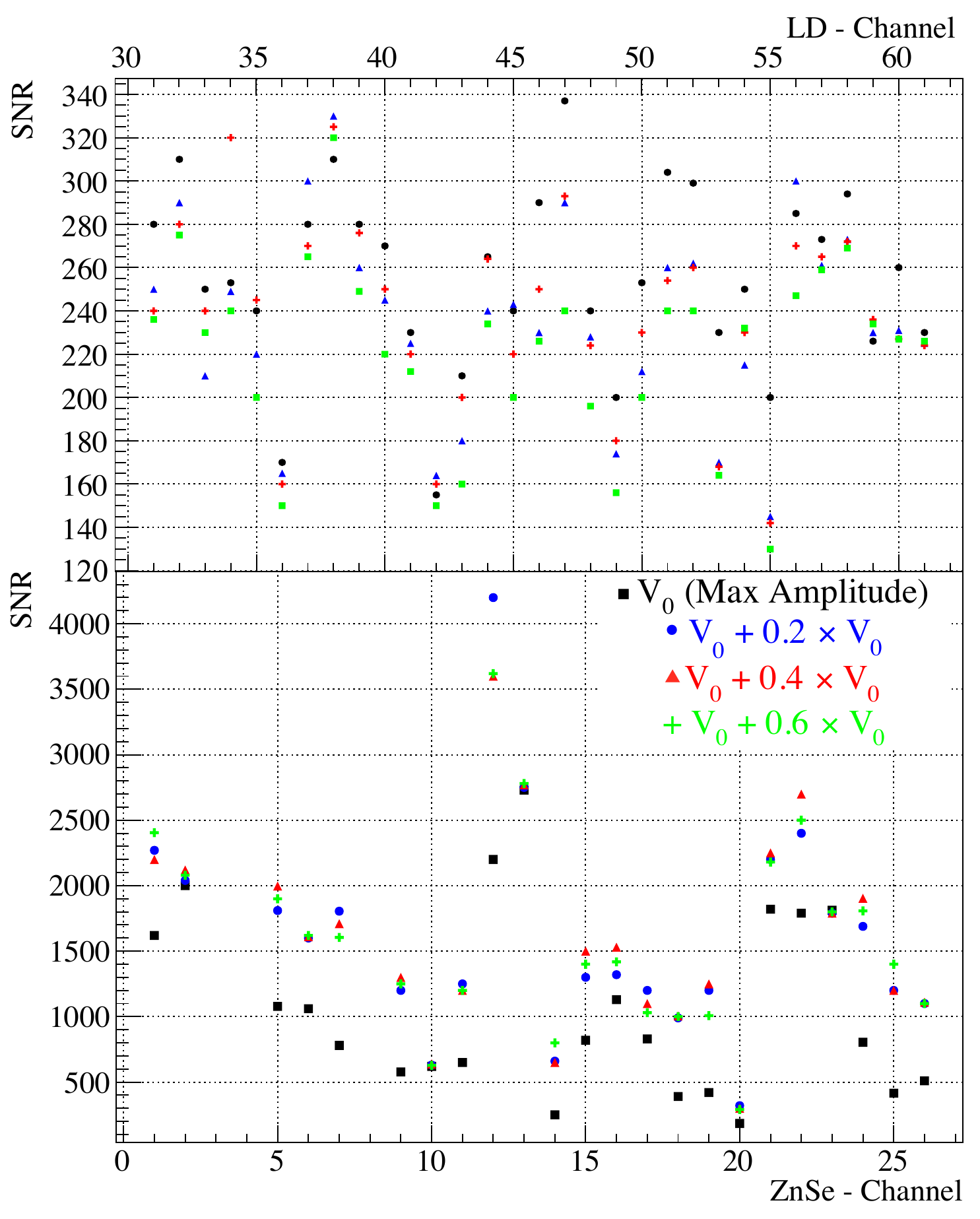}
\caption{Signal-to-noise ratio (SNR) scan for LDs (top) and ZnSe crystals (bottom) varying the voltage bias. It is calculated as the ratio of the filtered pulser amplitude to the $\sigma_{baseline}$. V$_0$(Max Amplitude) represents the biasing voltage which gives the maximal pulse amplitude. The scan is performed at bias higher than V$_0$(Max Amplitude) because we expect a stronger reduction of the noise compared to the signal amplitude.} 
\label{fig:snr}
\end{centering}
\end{figure}

In order to better estimate the best operating conditions of the detectors, the signal-to-noise ratio is the key parameter to be optimized. In fact, also the noise amplitude has to be taken into account, especially the parallel Johnson noise that develops across the resistors of the biasing circuit, which becomes more relevant at high values of the Ge-NTD resistances, hence lower temperatures. A compromise between low noise condition - higher temperature - and large signal amplitudes - lower temperature - must be established.
A reference pulse is generated on each detector dissipating the same amount of energy through the Si resistors. While varying the biasing voltage we monitor how the amplitude of the reference signal varies and how the detector noise changes. In Fig.~\ref{fig:snr} we show the signal-to-noise ratio for a set of measurements at different detector operating bias for ZnSe and LD's. The reported values are estimated filtering the acquired pulses by means of the Optimum Filter technique~\cite{OF}.

Summarizing, for each detector load curves measurements are performed for evaluating the configurations that maximize the signal amplitudes, see Fig.~\ref{fig:loadcurve_vi}. Then, a narrower scan in proximity of these defined working operation is carried out, aiming at defining the best signal-to-noise ratio for each detector, see Fig.~\ref{fig:snr}.

\section{Detector performance}

The overall detector performance are benchmarked by means of a $^{232}$Th calibration source deployed next to the detector, but outside of the cryostat. This is used to calibrate the energy response of the detector and to evaluate the detector energy resolution at the RoI but also the detector baseline energy resolution~\footnote{The energy resolution at 0~keV is evaluated as the detector baseline resolution and it is evaluated on acquired baseline where no pulses are recorded.}. Unfortunately, we are only able to calibrate the ZnSe crystals and not the LDs, due to their low mass. The best method to calibrate such small devices would be to place a permanent X-ray source on the detector, as it was already done in~\cite{Zn82Se}. In CUPID-0 we decided not to install any sort of permanent source on the detector for obvious reason related to the ultra-low background conditions in which the measurement is carried out.

\begin{figure}
  \begin{subfigure}[b]{0.49\columnwidth}
    \includegraphics[width=\columnwidth]{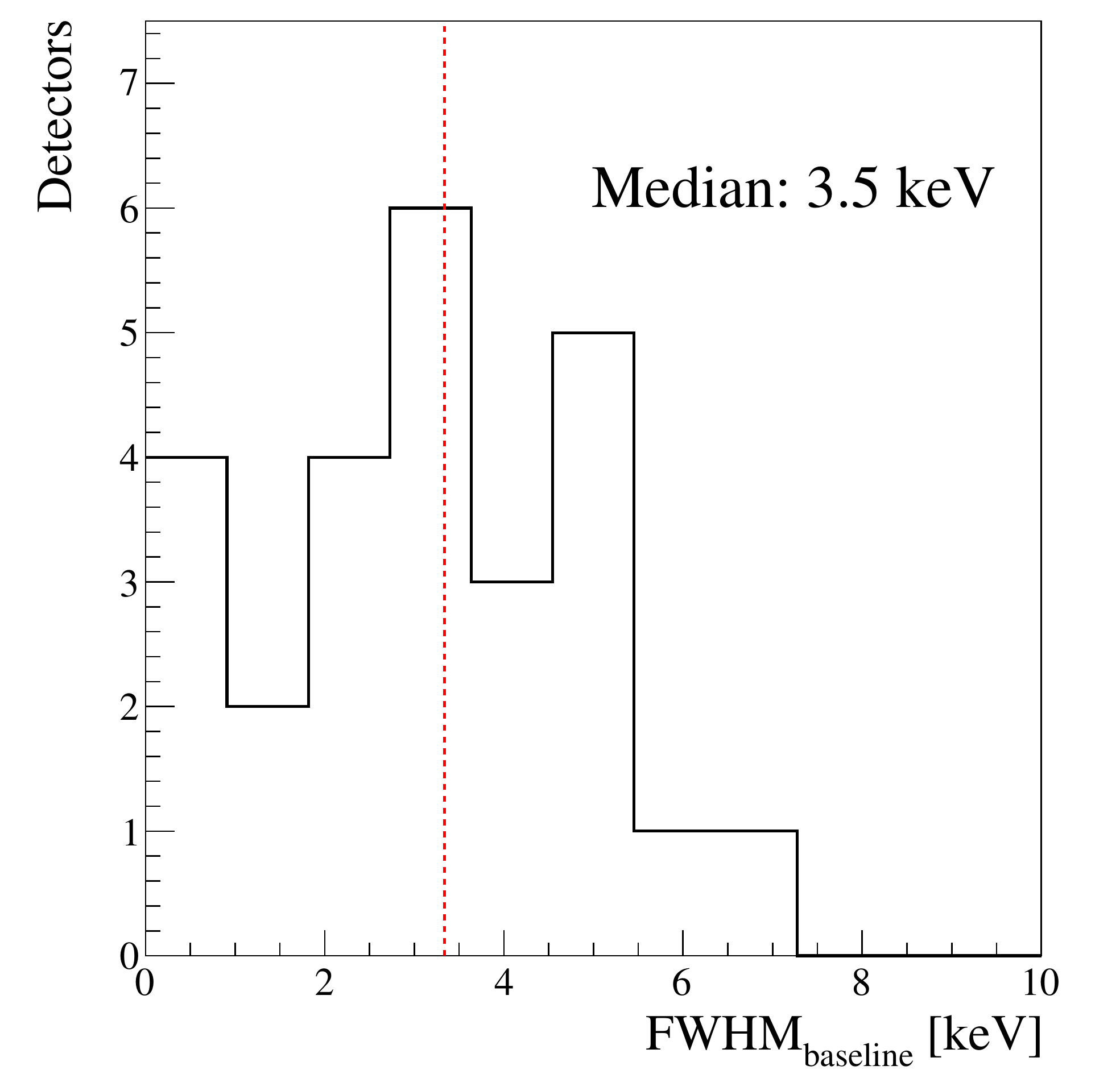}
    \caption{Energy resolution distribution at 0~keV for the ZnSe detectors.}
    \label{fig:0keV}
  \end{subfigure}
  \hfill
  \begin{subfigure}[b]{0.49\columnwidth}
    \includegraphics[width=\columnwidth]{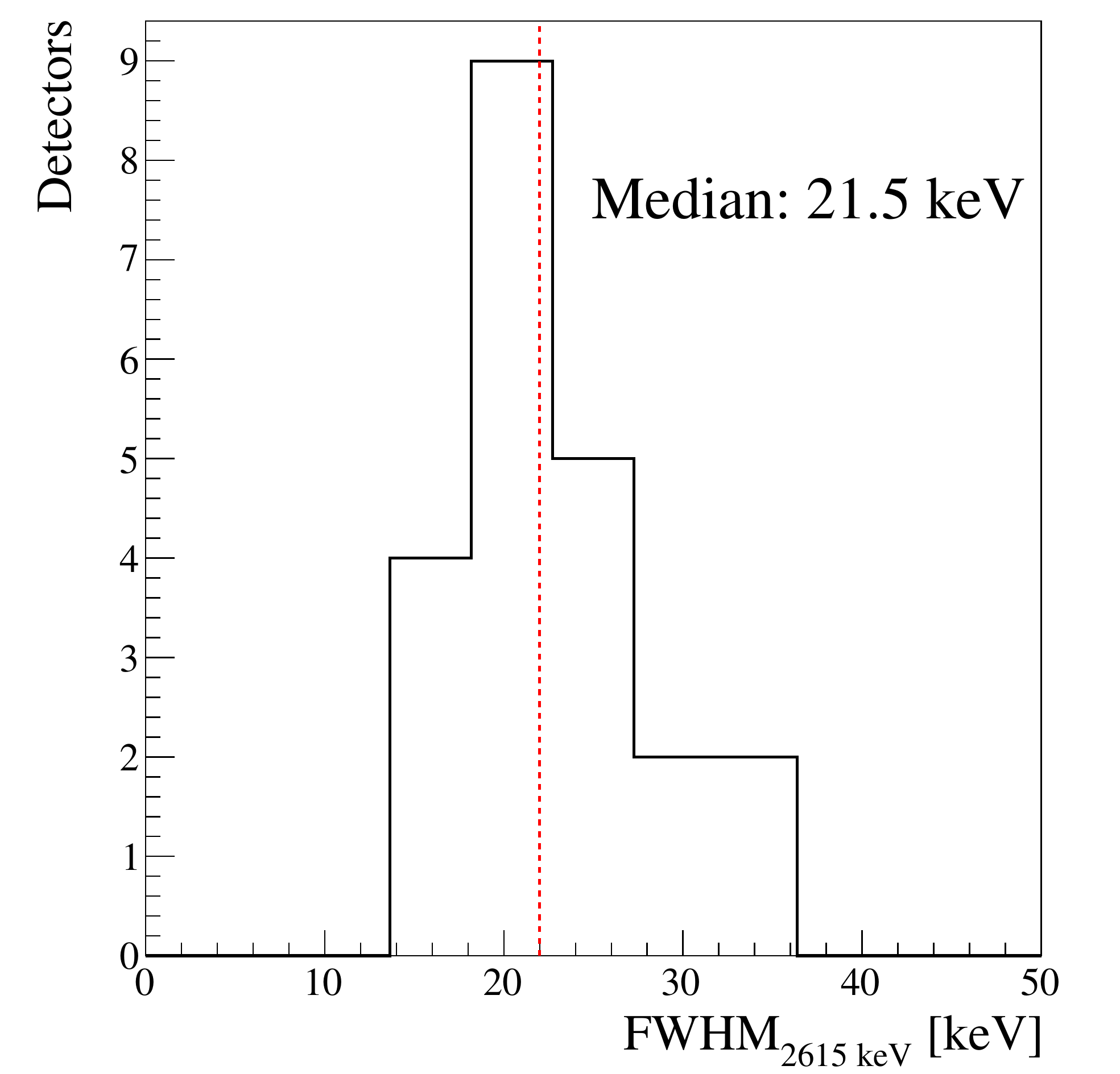}
    \caption{Energy resolution distribution at 2615~keV for the ZnSe detectors.}
    \label{fig:2615}
  \end{subfigure}
  \caption{Distribution of the ZnSe energy resolutions. On the left is shown the FHWM resolution at 0~keV, which is defined as the detector baseline noise. On the right the distribution for the 2615~keV $\gamma$-line energy is shown.}
\end{figure}

In Fig.~\ref{fig:0keV} and Fig.~\ref{fig:2615} we show the distribution of the detector FHWM resolutions for the ZnSe detectors at 0~keV (FWHM$_{baseline}$) and at 2.6~MeV (FWHM$_{2615}$), the high energy and high intensity $\gamma$-line produced by the $^{232}$Th source. The median value of the FWHM$_{baseline}$ reveals us that the cryogenic system and the electronics are performing at the cutting edge, given that the baseline noise in first approximation is independent of the absorber properties.  

The average detector energy resolution is computed at 2.6~MeV, the most intense high energy gamma line next to the region of interest. The exposure-weighted harmonic mean FWHM energy resolution results to be 23.0$\pm$0.6~keV. 
The spread in the energy resolution is driven by the limited crystal quality, in fact while an ideal bolometer is supposed to be a crystal with a single-crystalline structure, our detectors have polycrystalline structures. This characteristic strongly affects the thermalization of phonons inside the crystal, hence the detector response function.
We would like to underline the fact that there are still effective methods for improving the detector energy resolutions, and the most important one consists in taking advantage of the heat-light correlation in ZnSe crystals. In our group, while operating ZnSe bolometers, we were able to improve the detector energy resolution by a 25\%~\cite{ZnSe} by means of the heat-light de-correlation.

\begin{figure}
\begin{centering}
\includegraphics[width=0.7\columnwidth]{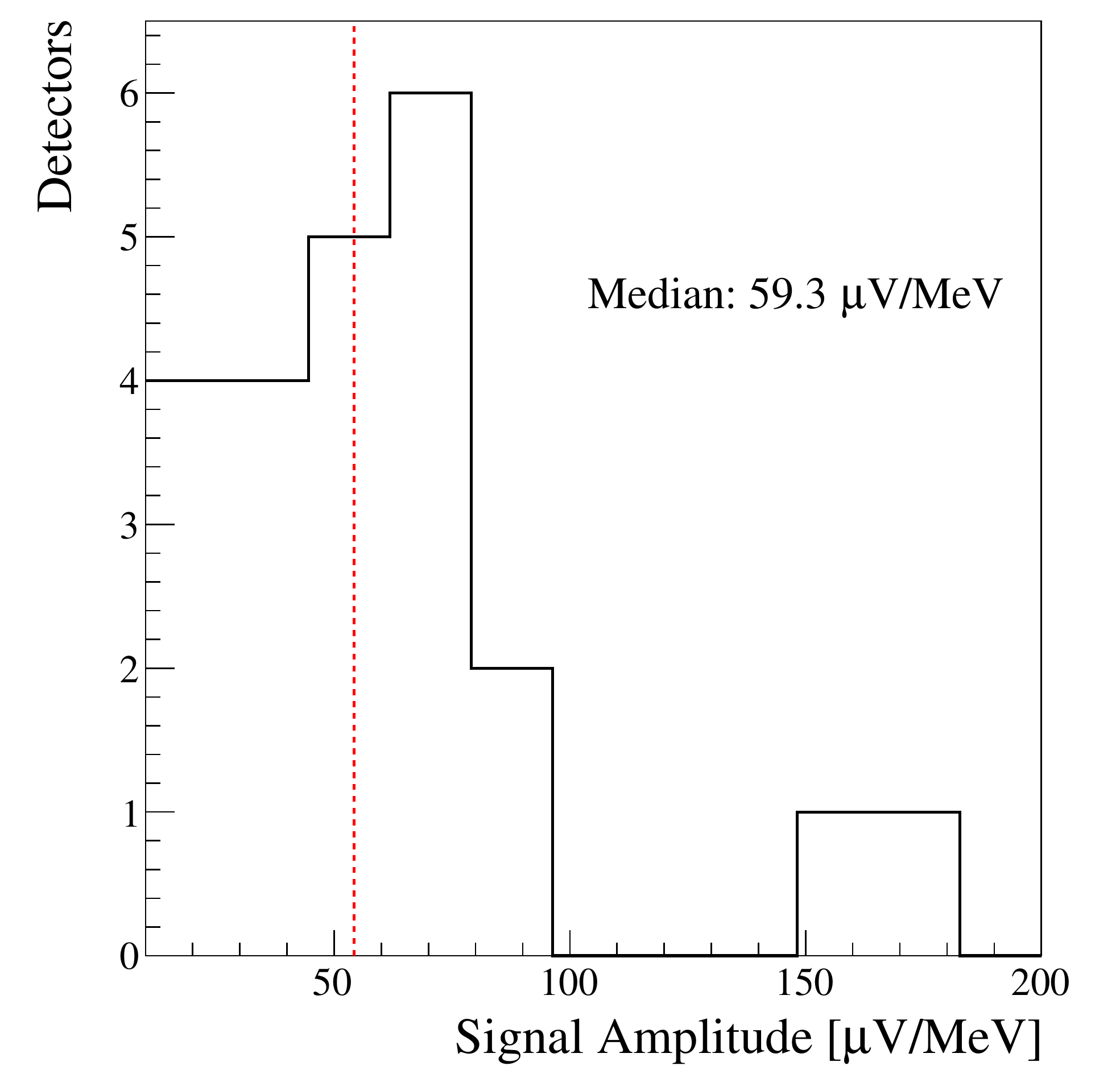}
\caption{Distribution of the ZnSe signal amplitudes for each crystal.}
\label{fig:SigAmpl_ZnSe}
\end{centering}
\end{figure}

In Fig.~\ref{fig:SigAmpl_ZnSe}, we show the distribution of the signal amplitudes for the Zn$^{82}$Se crystals. The median value of the distribution is 59.3~$\mu$V/MeV, which is a value comparable with other large mass bolometers like $^{nat}$TeO$_2$~\cite{CUORE-0_detector}.
 
\begin{figure}
\begin{centering}
\includegraphics[width=0.7\columnwidth]{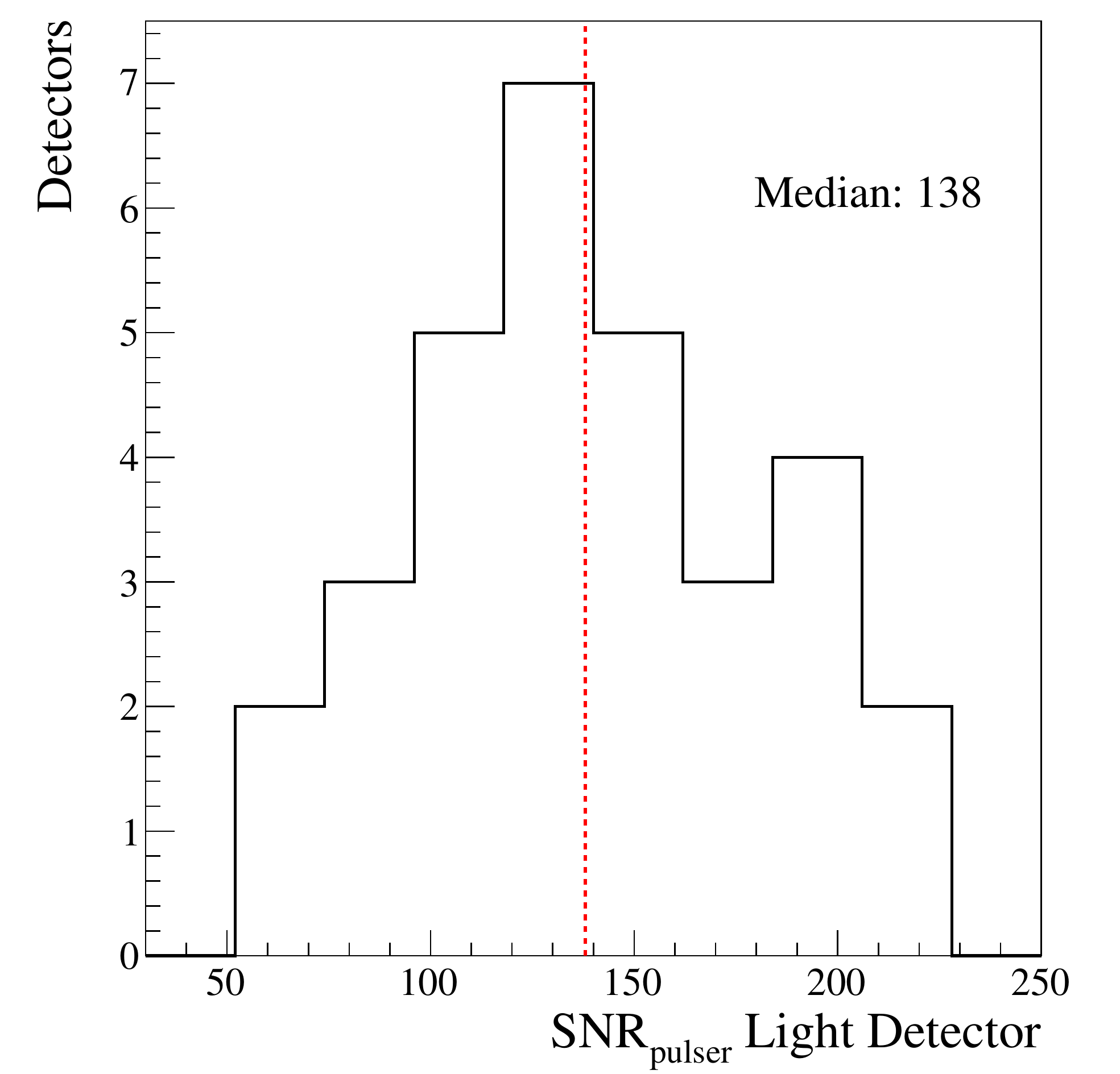}
\caption{Distribution of the LD SNRs. The signal amplitude is evaluated on test pulses generated by Si resistor coupled to each LD. The amount of dissipated energy is the same for each detector.}
\label{fig:LD_SNR}
\end{centering}
\end{figure}

This is the first time that a large number of LDs is operated: 31 channels. Their performance were never investigated in such a large scale. Due to the lack of a calibration source for the LD, we investigate the LD performance by means of the pulses generated by Si resistor coupled to the absorber. In Fig.~\ref{fig:LD_SNR}, the distribution of the signal-to-noise ratio of the heater pulses recorded on each LD is shown. The detector performance are extremely reproducible showing no major outlier.

Finally in Tab.~\ref{tab:per4}, we show the overall operating and performance parameters of the detectors. The rise-time and the decay time are also shown. These are computed as the time interval between the 10\% and 90\% of the leading edge of the pulse amplitude and as the 90\% and 30\% of the trailing part of the pulse amplitude, respectively. Furthermore we also report the noise amplitude evaluated at 5~Hz which is within the signal bandwidth. This shows that the detector resolution is limited not by the electronics noise (see Sec.~\ref{sec:readout}), but by the detector theirself.

We would like to underline the fact that the energy calibration of the LD it is not needed in order to perform the particle identification and rejection, because this is carried out on the relative signal amplitude. Moreover, in Tab.~\ref{tab:per4} given the reproducibility of the LD performance we only show the median value for the different parameters.

\begin{table*}[t]
\centering
\caption{Summary of the main operating parameters of the CUPID-0 detectors. For the LDs only the median values are reported, given the small spread in the performance. Three detectors (Channel ID 3, 4, 8) have a reduced signal amplitude which prevented us from evaluating the detector energy resolution. }
\resizebox{\textwidth}{!}{
\begin{tabular}{lccccccccccccc}
\hline
\hline
Channel ID&Name&Tower&Type&Mass&R$_{base}$&R$_{work}$&Noise@5Hz&Signal Amplitude&Rise time&Decay time&FWHM$_{baseline}$&FWHM$_{2615}$\\
&&&&[g]&[M$\Omega$]&[M$\Omega$]&[nV/$\sqrt{Hz}$]&[$\mu$V/MeV]&[ms]&[ms]&[keV]&[keV]\\
\hline 
1&CG-01&1&Enriched&439.40&10&4.17&11.7&40&10.1&24.7&2.59&21\\
2&CG-13&1&Enriched&427.86&54&14.71&25.7&81&12.0&37.6&2.58&35\\
3&CG-28&1&Enriched&427.00&29&9.63&18.1&-&9.5&17.3&-&-\\
4&NAT-1&1&Natural&418.39&34&11.28&26.3&-&5.5&20.7&-&-\\
5&CG-26&1&Enriched&408.22&7&3.94&20.9&51&11.1&33.1&3.22&19\\
\hline
6&CG-02	&2&Enriched&441.29&16&4.90&19.3&46&9.3&23.6&3.86&22\\
7&CG-15	&2&Enriched&469.64&25&11.25&17.2&47&12.6&26.2&2.88&20\\
8&CG-29	&2&Enriched&480.90&23&6.42&17.7&-&8.5&13.7&-&-\\
9&CG-14	&2&Enriched&470.59&27&9.56&26.5&66&10.3&21.5&3.57&20\\
10&CG-16&2&Enriched&260.52&2&1.22&10.1&13&9.2&26.4&6.09&19\\
\hline
11&CG-03&3&Enriched&438.65&11&4.34&13.7&23&10.3&27.4&4.84&22\\
12&CG-20&3&Enriched&214.62&14&6.42&21.0&179&18.1&48.3&0.96&15\\
13&CG-23&3&Enriched&174.89&33&9.85&25.7&156&12.0&34.2&1.47&14\\
14&Xtra-4&3&Enriched&409.88&63&12.69&23.6&26&11.5&26.7&6.85&25\\
15&CG-18&3&Enriched&410.31&77&18.13&29.0&57&14.5&33.3&3.17&24\\
16&CG-22&3&Enriched&418.92&65&21.26&50.7&68&15.1&47.5&3.84&18\\
\hline
17&CG-04&4&Enriched&442.32&150&30.78&46.1&69&15.8&40.4&4.10&25\\
18&Xtra-2&4&Enriched&442.43&101&23.63&62.9&73&14.4&35.7&5.05&25\\
19&CG-17&4&Enriched&474.22&73&20.76&32.1&36&14.0&36.6&5.11&29\\
20&NAT-2&4&Natural&431.21&216&41.40&63.7&38&15.9&32.9&8.41&38\\
21&CG-21&4&Enriched&233.08&23&10.06&27.2&61&14.6&58.1&2.40&20\\
\hline
22&CG-10&5&Enriched&440.47&14&6.91&17.2&68&14.4&39.3&1.94&19\\
23&CG-08&5&Enriched&431.00&79&28.19&53.2&84&18.1&44.0&3.36&17\\
24&CG-24&5&Enriched&429.62&42&14.33&26.7&62&11.9&26.1&3.06&29\\
25&CG-25&5&Enriched&434.51&14&5.32&18.3&35&9.7&20.5&4.84&26\\
26&CG-27&5&Enriched&431.18&11&5.77&13.7&21&10.6&26.8&5.36&25\\
\hline
ZnSe Median&&&&&28&9.95&24.6&59.3&13.5&35.7&3.47&22\\
\hline
LD Median&&&&&6.1&4.16&6.5&-&3.5&7.1&-&-\\ 
\hline
\hline
\end{tabular}
}
\label{tab:per4}
\end{table*}

\section{Detector radiopurity}
In order to achieve the extremely low-background index in the RoI for a sensitive investigation of $0\nu\beta\beta$ decay, the detector radiopurity is fundamental. All the materials used for the detector were chosen for their ultra-low concentration of impurities. Nevertheless, the final detector radiopurity can be spoiled if dedicated procedures are not adopted, while producing or handling detector components. In order to validate and to prove the firm control of all the procedures adopted for the detector production, it worths to analyse the internal contamination of all the crystals used for CUPID-0. At the same time the study of the internal contaminations of the detector is fundamental for the development of a reliable and robust background model for the study of the possible background sources in the RoI.

\begin{figure}
\begin{centering}
\includegraphics[width=1.0\columnwidth]{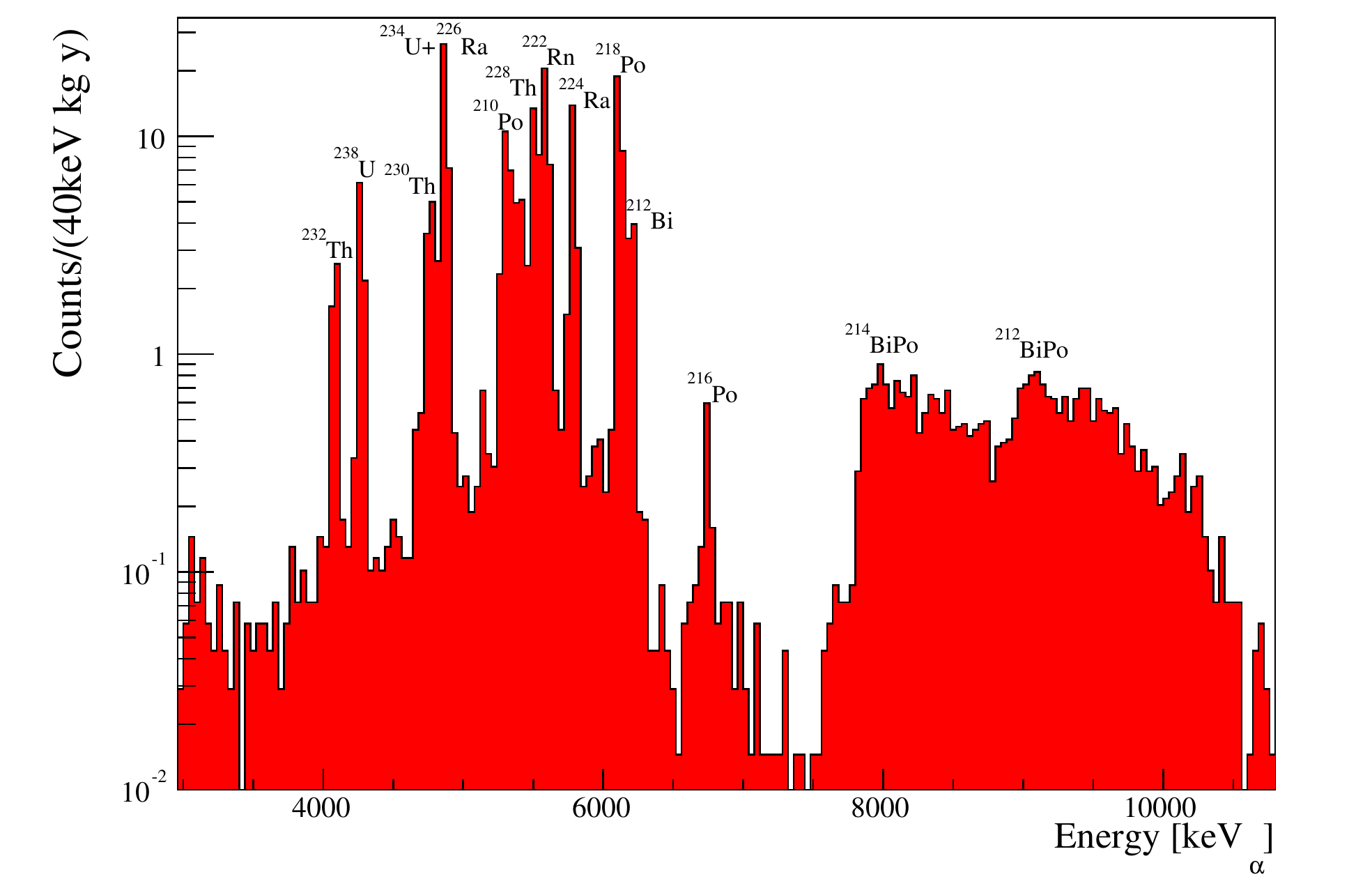}
\caption{Total alpha energy spectrum of all detectors.}
\label{fig:alpha}
\end{centering}
\end{figure}

Thanks to the excellent particle discrimination, thoroughly discussed in~\cite{Zn82Se,ZnSe}, we can select with high efficiency only events induced by $\alpha$ particle interactions. In Fig.~\ref{fig:alpha} we show the energy spectrum of the CUPID-0 detector for $\alpha$ interacting particles.

In the energy spectrum, the peaks between 4~MeV and 7~MeV are induced by natural radioactive decays occurring in the crystal bulk, while the excess of events at higher energies are induced by pile-up events, usually defined as Bi-Po cascade. In the $^{238}$U and $^{232}$Th decay chains there are two nuclides which undergo a $\beta$ decay and after few msec under an $\alpha$ decay. The poor time resolution of bolometer does not allow to disentangle these two events which then produce a single events with a total energy equal to the Q-value of the two transition minus the energy carried away by the neutrino. These cascades are $^{214}$Bi$\rightarrow$$^{214}$Po for the $^{238}$U chain and  $^{212}$Bi$\rightarrow$$^{212}$Po for the $^{232}$Th one.

All the other events in the energy spectrum are mostly ascribed to surface $\alpha$ contaminations~\cite{sticking}.

In Tab.~\ref{tab:cont} we report the measured internal radioactive contaminations for the CUPID-0 enriched crystals. We also show the highest and lowest concentrations observed in each nuclide. The overall detector radiopurity complies with the needs for a extremely low background index in the RoI of $^{82}$Se, which is expected to be at the level of few 10$^{-3}$~counts/(keV$\cdot$kg$\cdot$y). The average detector purity is not better than other bolometric detectors for $0\nu\beta\beta$~\cite{TeO2production}, nevertheless if we look at the best values for each nuclide, the achieved radiopurity is competitive with the previously mentioned results.

An important information that can be inferred from the table is the wide spread in the radiopurity level. In fact there is about one order of magnitude difference between the lowest and highest nuclide activity. All the procedure established for the crystal production were not sufficient for keeping under a complete control the impurity concentration in the crystals. Nevertheless, from preliminary studies we can state that the crystal purity has strongly improved during the crystal production, the crystals which show the highest purity are the one produced at end of the production campaign. A speculative explanation could be given by the fact that the crucibles employed for the crystal growth have not undergone a full purification process, which actually occurred during the crystal production.

\begin{table}[htp]
\caption{Evaluated internal $\alpha$ radioactive contamination for the CUPID-0 detector. The values reported refer to the overall detector radiopurity (CUPID-0), the lowest (Best) and highest (Worst) measured contamination in a single crystal. $^{210}$Po values refers to the sum of bulk and surface contaminations. Limits are evaluated at 90\% C.L..} 
\begin{center}
\begin{tabular}{lcccc}
\hline\noalign{\smallskip}
Chain & Nuclide  & Activity & Activity &  Activity\\ 
            & & [$\mu$Bq/kg] & [$\mu$Bq/kg] & [$\mu$Bq/kg]\\
\noalign{\smallskip}\hline\noalign{\smallskip}
  & & CUPID-0 & Lowest & Highest \\
\hline
$^{232}$Th & & & &\\
 & $^{232}$Th & 2.5$\pm$0.2 & $<$0.54  & 8.6$\pm$1.2\\
  & $^{228}$Th & 13.6$\pm$0.4 & 2.3$\pm$0.8 & 26.9$\pm$2.2\\
   & $^{224}$Ra & 10.9$\pm$0.3 &  2.1$\pm$0.6 & 23.1$\pm$0.2 \\
   & $^{212}$Bi & 12.2$\pm$0.6 & $<$3.7 & 24.2$\pm$3.5 \\
\noalign{\smallskip}\hline\noalign{\smallskip}
$^{238}$U & & & &\\
 & $^{238}$U & 5.1$\pm$0.2 & $<$1.2 & 12.7$\pm$1.5  \\
 & $^{234}$U & 5.3$\pm$0.8 & 1.0$\pm$2.0 & 14.7$\pm$4.3 \\ 
 & $^{230}$Th & 5.3$\pm$0.2 & $<$2.4 &16.4$\pm$1.7 \\
 & $^{226}$Ra & 17.0$\pm$0.4 & 3.8$\pm$0.9 & 18.4$\pm$1.8\\
 & $^{218}$Po & 17.4$\pm$0.4 & 3.4$\pm$0.6 & 19.8$\pm$1.9\\
 & $^{210}$Po & 18.8$\pm$0.6 & 9.1$\pm$0.3 & 45.4$\pm$2.6\\
\noalign{\smallskip}\hline\noalign{\smallskip}
\end{tabular}
\label{tab:cont} 
\end{center}
\end{table}

\section{Conclusion}
CUPID-0 is the first large array of scintillating bolometers for $0\nu\beta\beta$ decay investigations. The detector is made of 26 crystals, 24 of them are enriched in $^{82}$Se at the level of 95\% and 2 have natural isotopic abundance. The large $0\nu\beta\beta$ source mass and the particle discrimination capability enable to reach unprecedented level of background in the RoI, never achieved with bolometers. This is possible thanks to the excellent performance and reproducibility of the LDs which are operated together with the Zn$^{82}$Se bolometers.

The large number of detectors, 31 LDs and 26 Zn$^{82}$Se crystals, required an upgrade of the cryogenic infrastructure. The cryostat is now able to host up to 136 channels, thus also a possible second phase of the CUPID program. Furthermore an innovative and effective vibration damping system was installed which allowed to achieve relevant results on the detector performance, in terms of baseline energy resolution.

The state of the art technologies in low background techniques and in cryogenic detector design have been implemented inside the CUPID-0 detector. The preliminary results obtained with this innovative detector are demonstrating the feasibility of a next generation cryogenic experiment for the investigation of $0\nu\beta\beta$ down to the meV scale.

\begin{acknowledgements}
This work was partially supported by the European Research Council (FP7/2007-2013) under contract LUCIFER no. 247115. We are particularly grateful to M. Iannone for the help in all the stages of the detector construction, A. Pelosi for construction of the assembly line, M. Guetti for the assistance in the cryogenic operations, M. Lindozzi for the development of cryostat monitoring system and the mechanical workshop of LNGS (in particular E. Tatananni, A. Rotilio, A. Corsi, and B. Romualdi) for continuous and constructive help in the overall set-up design. We acknowledge the Dark Side collaboration for the use of the low-radon clean-room.
\end{acknowledgements}

\end{document}